\documentclass{WileyMSP-template}
\usepackage{caption}
\usepackage{textcomp}
\usepackage{subcaption}
\usepackage{amsmath}
\usepackage{setspace}
\usepackage{siunitx}
\usepackage{cite}
\begin{document}


\rhead{\includegraphics[width=2.5cm]{vch-logo.png}}

\title{Impact of Side Chain Hydrophilicity on Packing, Swelling and Ion Interactions in Oxy-bithiophene Semiconductors}

\maketitle


\author{Nicholas Siemons*}
\author{Drew  Pearce}
\author{Camila Cendra}
\author{Hang Yu}
\author{Sachetan M. Tuladhar}
\author{Rawad K. Hallani}
\author{Rajendar Sheelamanthula}
\author{Garrett S. LeCroy}
\author{Lucas Siemons}
\author{Andrew J. P. White}
\author{Iain Mcculloch}
\author{Alberto Salleo}
\author{Jarvist M. Frost}
\author{Alexander Giovannitti}
\author{Jenny Nelson*}



\begin{affiliations}

\vspace{0.3cm}

N. Siemons, \\
Department of Physics,  \\
Imperial College, London, \\
Exhibition Rd, South Kensington, London, SW7 2AZ, UK, \\
Email Address: n.siemons18@imperial.ac.uk \\

\vspace{0.3cm}

Dr. D. Pearce, \\
Department of Physics,  \\
Imperial College, London, \\
Exhibition Rd, South Kensington, London, SW7 2AZ, UK \\

\vspace{0.3cm}

Dr. C. Cendra, \\
Department of Materials Science and Engineering, \\
Stanford University,  \\
450 Serra Mall, Stanford, CA 94305, United States \\
 
\vspace{0.3cm}

H. Yu, \\
Department of Physics,  \\
Imperial College, London, \\
Exhibition Rd, South Kensington, London, SW7 2AZ, UK \\

\vspace{0.3cm}

Dr. S. M. Tuladhar, \\
Department of Physics,  \\
Imperial College, London, \\
Exhibition Rd, South Kensington, London, SW7 2AZ, UK \\

\vspace{0.3cm}

Dr. R. K. Hallani, \\
KAUST Solar Center,  \\
King Abdullah University of Science and Technology (KAUST), \\
Thuwal 23955, Saudi Arabia \\

\vspace{0.3cm}

Dr. R. Sheelamanthula, \\
KAUST Solar Center,  \\
King Abdullah University of Science and Technology (KAUST), \\
Thuwal 23955, Saudi Arabia \\

\vspace{0.3cm}

G. S. LeCroy, \\
Department of Materials Science and Engineering, \\
Stanford University,  \\
450 Serra Mall, Stanford, CA 94305, United States \\

\vspace{0.3cm}

Dr. L. Siemons, \\
The Francis Crick Institute, \\
1 Midland Road, London, NW1 1AT, UK \\

\vspace{0.3cm}

Dr. A. J. P. White \\
Chemical Crystallography Laboratory, Department of Chemistry, \\
Imperial College London White City Campus, 82 Wood Lane, London, UK, W12 0BZ \\

\vspace{0.3cm}

Prof. I. Mcculloch, \\
Department of Chemistry, \\
University of Oxford, Oxford, OX1 2JD, UK \\

\vspace{0.3cm}

Prof. A. Salleo, \\
Department of Materials Science and Engineering, \\
Stanford University,  \\
450 Serra Mall, Stanford, CA 94305, United States \\

\vspace{0.3cm}

Dr. J. M. Frost \\
Department of Physics,  \\
Imperial College, London, \\
Exhibition Rd, South Kensington, London, SW7 2AZ, UK \\

\vspace{0.3cm}

Dr. A. Giovannitti, \\
Department of Materials Science and Engineering, \\
Stanford University,  \\
450 Serra Mall, Stanford, CA 94305, United States \\

\vspace{0.3cm}

Prof. J. Nelson \\
Department of Physics,  \\
Imperial College, London, \\
Exhibition Rd, South Kensington, London, SW7 2AZ, UK \\

\end{affiliations}

\vspace{0.3cm}



\begin{abstract}
Exchanging hydrophobic alkyl-based side chains to hydrophilic glycol-based side chains is a widely adopted method for improving mixed-transport device performance, despite the impact on solid state packing and polymer-electrolyte interactions being poorly understood. Presented here is a Molecular Dynamics (MD) force field for modelling alkoxylated and glycolated polythiophenes.  The force field is validated against known packing motifs for their monomer crystals.  MD simulations, coupled with X-ray Diffraction (XRD), show that alkoxylated polythiophenes will pack with a `tilted stack' and straight interdigitating side chains, whilst their glycolated counterpart will pack with a `deflected stack' and an s-bend side chain configuration.  MD simulations reveal water penetration pathways into the alkoxylated and glycolated crystals - through the  $\pi$-stack and through the lamellar stack respectively. Finally, the two distinct ways tri-ethylene glycol polymers can bind to cations are revealed, showing the formation of a meta-stable single bound state, or an energetically deep double bound state, both with a strong side chain length dependance.  The minimum energy pathways for the formation of the chelates are identified, showing the physical process through which cations can bind to one or two side chains of a glycolated polythiophene, with consequences for ion transport in bithiophene semiconductors. 
\end{abstract}

\vspace{0.5cm}
Mixed-ionic-electronic conducting conjugated polymers (OMIECs) underpin the operation of a range of bioelectronic devices \cite{rivnay2018organic, owens2010interface, moser2019materials}, such as organic electrochemical transistors (OECTs) \cite{savagian2018balancing, Giovannitti2018, Giovannitti2016}, batteries \cite{Moia2019} and super capacitors \cite{Giovannitti2016}. They have inherent advantages over traditional organic semiconductors for these electrochemical applications \cite{Rivnay2016,Nielsen2016, wang2019hybrid, savagian2018balancing}, in particular their ability to operate with an aqueous electrolyte \cite{pappa2018direct, ohayon2019biofuel}.  Exchanging a traditional alkyl-based side chain for a glycol-based hydrophilic side chain has been widely adopted as a strategy for increasing uptake of water and ions in OMIECs, and subsequently increasing their capacitance \cite{kroon2017polar, Moia2019, kiefer2018enhanced, Giovannitti2018, chen2021n}.  Whilst alkylated conjugated polymers are known to generally pack with interdigitating side chains and straight $\pi$-stacks \cite{tashiro1997crystal, Winokur1991, hugger2004semicrystalline, brinkmann2006orientation, wu2010temperature, dudenko2012strategy} (see Figure S1 for backbone packing motifs referred to in this study), little is known about glycolated OMIEC packing, despite the chain packing being critical to both electronic and ionic transport \cite{coropceanu2007charge, Tseng2014, Flagg2019, kim2018influence, mcculloch2006liquid}. 

\vspace{0.4cm}
Experimental studies have suggested that glycolated OMIECs adopt smaller $\pi-$stack distances than their alkylated counterparts \cite{Meng2015, Giovannitti2018}. As well as solid-state packing, structural characterisation of OMIECs should account for their swelling behaviour, since electrochemical doping is known to cause volumetric swelling in these systems \cite{Flagg2019, Savva2020}.  XRD studies have thus correlated doping with an increase in lamellar stack spacing for both alkylated and glycolated polymers \cite{Kawai1992, Guardado2017, Thelen2015, Thomas2018, Savva2019, Cendra2019}, implying electrolyte entry into the lamellar stack. However, as the majority of the polymer is non-crystalline, X-ray diffraction (XRD) alone is unable to characterise many features critical to charge transport \cite{Baumeier2010, Valeev2006, Nelson2009, winkler1999electron, page1999natural}.  Furthermore, it is unable to determine the position of water and ions in the lattice beyond inferring it from the polymer structure.  To obtain this level of detail it is necessary to use experimental methods such as solid state nuclear magnetic resonance (ssNMR) crystallography \cite{dudenko2012strategy, taulelle2004nmr, harris2009nmr}, or computational methods such as MD. 

\vspace{0.4cm}
MD offers the means to simulate atomistic structure and dynamics for both ordered and amorphous polymers, as well as for liquids, but relies on careful force field validation \cite{Jorgensen1996, wang2004development, stone1996theory}, especially for $\pi-$conjugated systems \cite{Moreno2010, Bhatta2013}.  Even with a validated force field,  finding stable polymer crystal motifs with MD is challenging due to its limited sampling of the morphological phase space \cite{giberti2015insight, giberti2015metadynamics, liu2020free, piaggi2018predicting, zykova2008investigating}. Therefore MD studies on the ordered phase of OMIECs have not attempted to do so, but ensured long range order through use restraining potentials and small simulation cells \cite{liu2021amphipathic}, or constraining systems to two dimensions \cite{ghosh2020effect}.  Some studies \cite{Gladisch2019, moser2021controlling, ghosh2020effect} have taken advantage of MDs ability to probe amorphous polymer phases to uncover phase transitions during doping, however, they use the General AMBER Force Field \cite{wang2004development} that has not been validated for the systems under study and is primarily designed for modelling of non-conjugated systems \cite{dubay2012accurate}. Other studies \cite{moser2020ethylene, matta2020ion} use the same force field, but reparameterise backbone dihedrals to accurately capture backbone torsional behaviour, a necessary step to accurately model the behaviour of conjugated polymers \cite{dubay2012accurate, wildman2016general, sundaram2020development}.  Some of the latter work \cite{matta2020ion}, as well as studies in closely related fields using more advanced validated force fields \cite{Savoie2017, khot2021side}, have made progress on understanding electrolyte structure in polymer films, showing that hydrophilic side chains will coordinate with cations, with implications for the operation of both hole and electron transporting OMIECs \cite{wang2019real, Flagg2020, zhou1987incorporation, chao1987ionic, paulsen2020organic}.  MD studies to date however have not been used to calculate water placement in an ordered polymer lattice.  Whilst MD studies have clearly shown the ability of cations to coordinate with glycol side chains,  free-energetics have been inferred from populations obtained from a limited sample of the morphology.  For MD studies of binding events in biological systems \cite{limongelli2013funnel, moraca2017ligand, troussicot2015funnel, barducci2006metadynamics, granata2013characterization, clark2016prediction, saleh2017efficient}, it is generally accepted that enhanced sampling methods such as metadynamics \cite{tiwary2016review, bernardi2015enhanced, paquet2015molecular, elber1996novel} must be used to sufficiently sample the morphological phase space, and we propose the same is true in polymer systems. 

\begin{figure}[b!]
\center
  \includegraphics[width=18cm]{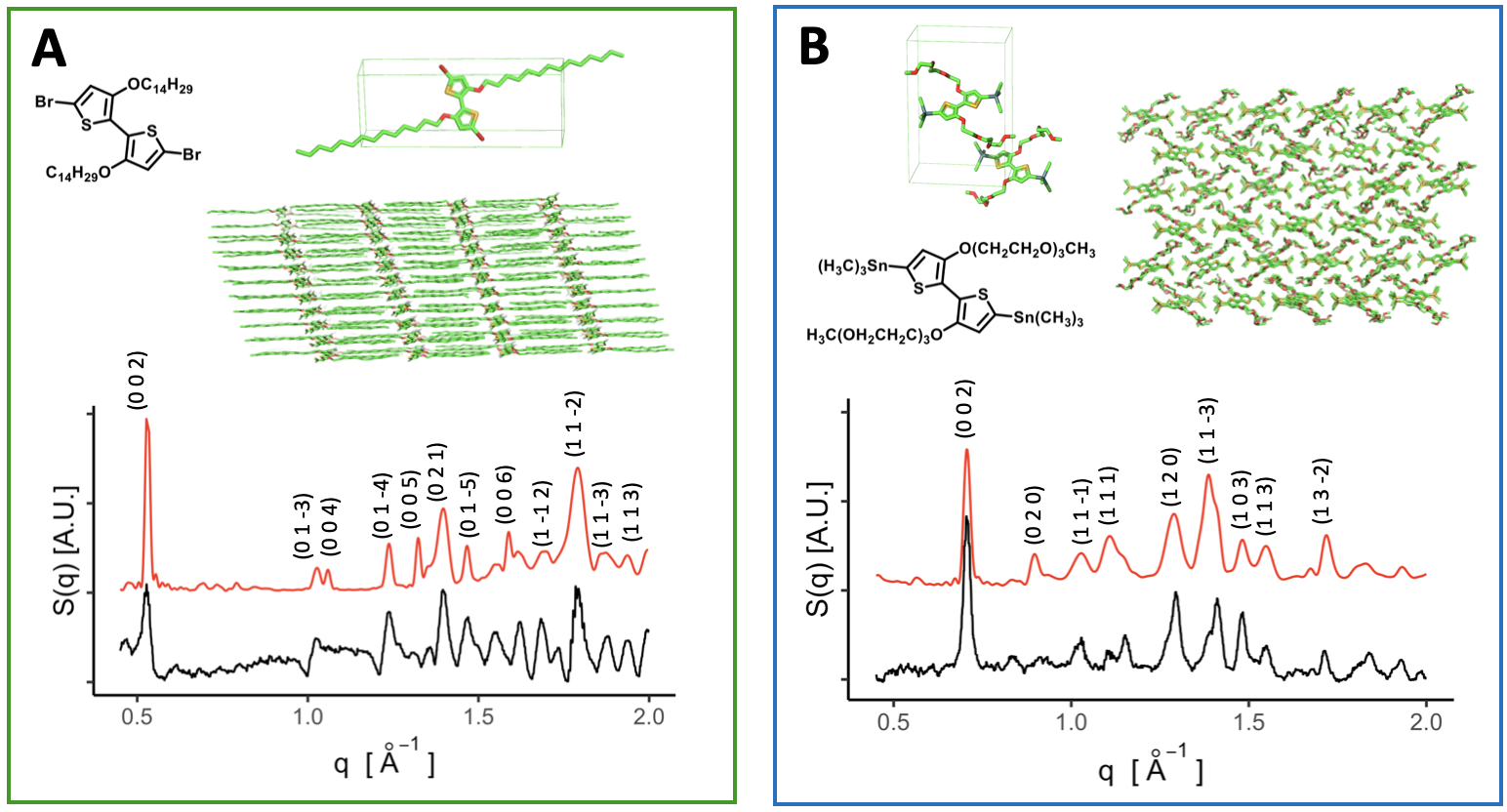}
  \caption{ Monomer crystal determination through experimentation, and MD force field validation with A) aT2 and B) gT2.  Each panel contains the chemical display formula of the molecule, an illustration of the unit cell and supercell, and the experimentally determined x-ray diffraction pattern (black line), as well as the simulated x-ray pattern from MD (red line).   Peak indices are allocated by comparison with the theoretical structure factors (Figure S16). Unit cell parameters are $a=4.17\,\si{\angstrom}$, $b=9.30\,\si{\angstrom}$, $c=23.80\,\si{\angstrom}$, $\alpha=91.8^o$, $\beta=91.5^o$, $\gamma=96.5^o$, P1 symmetry for aT2 and $a=7.00\,\si{\angstrom}$, $b=14.02\,\si{\angstrom}$, $c=18.01\,\si{\angstrom}$, $\alpha=90.00^o$, $\beta=97.71^o$, $\gamma=90.00^o$, P21/n symmetry for gT2. }
\end{figure}



\vspace{0.4cm}
To understand the packing behaviours of glycolated OMIECs, we have taken the archetypal glycolated polythiophene, poly(3,3’dialkoxy‑(triethyleneglycol)-bithiophene) (p(gT2)) \cite{Moia2019}, and synthesised an alkylated analogue for comparison, poly(3,3’ditetradecoxy‑bithiophene) (p(aT2)). Due to the use of an oxygen atom at the side chain attachment point, these species will be referred to as having `alkoxy' side chains. Use of alkoxy side chains ensures the backbone behaviours are analogous, as the oxygen at the side chain attachment point is known to strongly influence the backbone morphology \cite{thorley2018s}. 

\vspace{0.4cm}
To further expand our understanding of packing behaviours in these systems, monomer crystals of these polymers are synthesised, aT2 and gT2, and the crystal packing motifs are atomistically characterised using XRD.  Using the monomer motifs,  we validate a force field for Molecular Dynamics simulations (see SI Section 2) which, when coupled with XRD measurements, is able to determine polymer packing motifs for these systems.  We proceed to investigate how the interaction with water, as well as water placement in the lattice, depend on the chemistry of the side chain. Finally we observe and quantitatively characterise the ion-side chain interactions that are unique to glycol side chains that have been intuitively known and only qualitatively studied thus far. 


\vspace{0.4cm}
Force field validation is achieved through close agreement between simulated and experimental XRD patterns for the monomer crystals, shown in \textbf{Figure 1}. The simulated supercells contain between 250 and 300  monomers and are stable throughout the simulation.  They further show stability when annealed at 350K, with the aT2 crystal also stable after annealing at 400K (Figure S9), corroborating the experimental observation that aT2 crystals are easier to grow than gT2 crystals. We see close agreement between experimental and simulated powder patterns, with almost all the peaks present in experiment being present in our simulated pattern. Additionally, relative peak intensities and widths are well reproduced, showing not only is the structure maintained, but also the relative levels of disorder between the crystallographic planes are reproduced. 

\vspace{0.4cm}
Utilising packing behaviours observed in the monomer crystals, we are able to hypothesise and test, in terms of stability, candidate crystal motifs for p(aT2) and p(gT2). For each candidate structure, a supercell of around 60 oligomers, each of 20 repeat units, is simulated for 50~ns using our validated force field.  For the structures that maintained their long-range order, simulated XRD powder patterns are calculated and compared with experiment.  After testing a variety of potential structures for both alkoxylated and glycolated crystals,  those shown in {\bf Figure 2} both show stability in our simulations,  and closely reproduce the experimentally measured XRD patterns.

\vspace{0.4cm}
We see close agreement between simulated and experimental polymer XRD patterns for the shown unit cells, with simulation showing additional peaks associated with side chain order.  Due to some of these peaks not arising in experiment, and close agreement between experiment and simulated patterns where side chain contributions are omitted (see Figure S17), we conclude that the crystalline polymer phase shows some level of disorder amongst the side chains.  Nevertheless, fitting the experimental peak widths to the peak order in the lamellar direction (see Figure S14) shows coherence over 4 and 8 lamellar planes for p(aT2) and p(gT2) respectively, indicating sufficient side chain ordering to maintain registration over multiple lamellae. Therefore peaks in experimental patterns, particularly at high $\vec{q}$ values, may still arise from side chains, or have additional intensity due to side chain scattering.

\vspace{0.4cm}
The alkoxylated monomer and polymer crystals share similarities, showing the same symmetry elements and side chain structure. The $\pi$-stacking structure is similar, with both crystals adopting a `tilted stack', and both crystals having straight interdigitating side chains.    Stronger differences are observed between the glycolated monomer and polymer crystals. In the monomer, the side chains adopt a curled structure and no $\pi$-stacking is observed.  For p(gT2),  the side chains adopt a gentle s-bend configuration, and a the backbones adopt a `deflected stack'.  In both cases however, the side chains arrange themselves so that oxygen atoms are neighboured by hydrogen atoms on other side chains. 

\begin{figure}[b!]
\center
  \includegraphics[width=19cm]{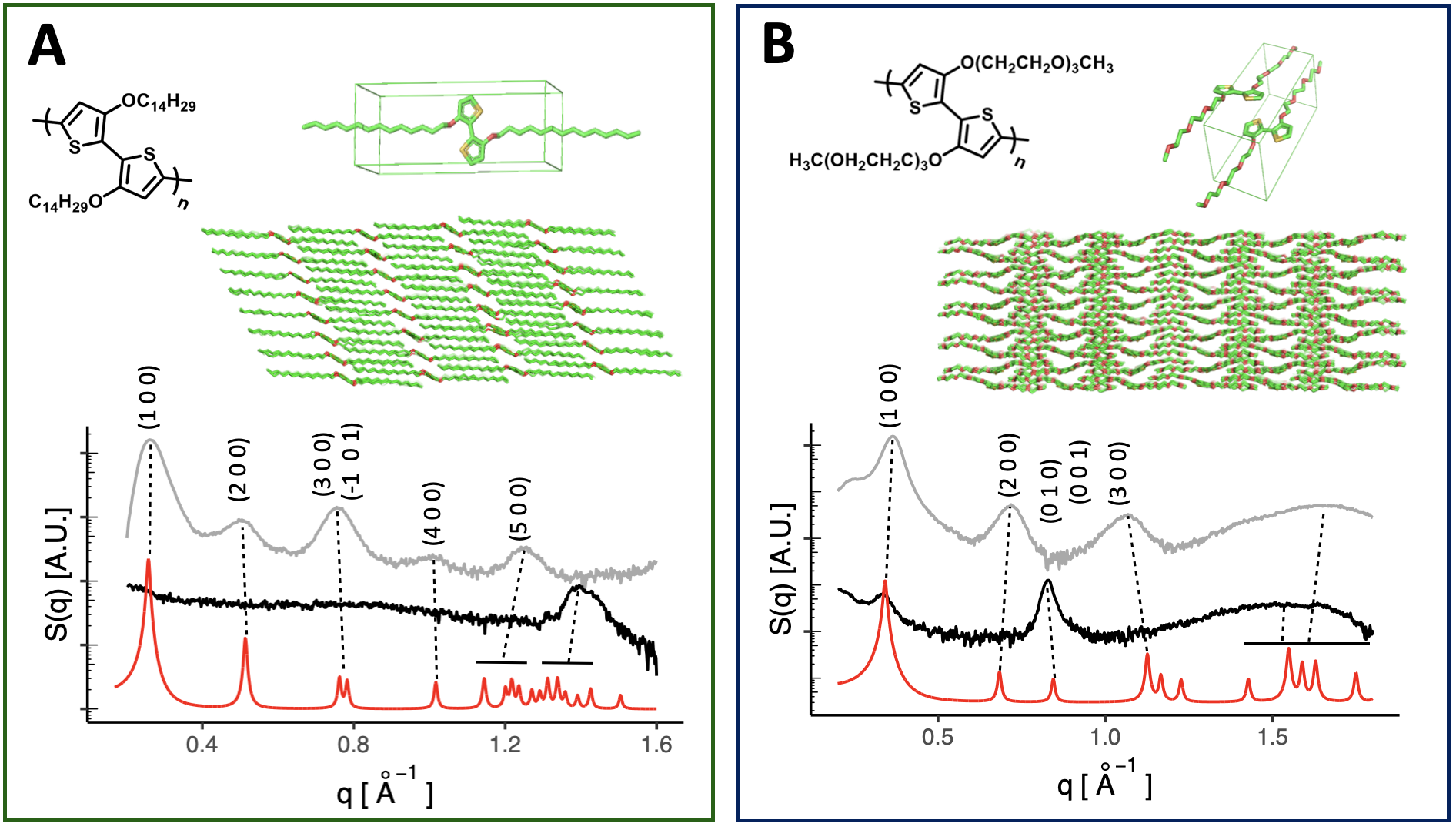}
 \caption{ Polymer crystal determination for A) p(aT2) and B) p(gT2).  Each panel contains the chemical display formula, an illustration of the unit cell and supercell, XRD patterns obtained from MD (red), experimental in-plane pattern (black) and experimental out-of-plane pattern (grey). See Figure S13 for 2D spectra. Peak indices are allocated by comparison with theoretical structure factors (Figure S16 and S17). Unit cell parameters are $a=22.4\,\si{\angstrom}$, $b=5.2\,\si{\angstrom}$, $c=7.4\,\si{\angstrom}$, $\alpha=69^o$, $\beta=86^o$, $\gamma=87^o$, P1 symmetry for p(aT2) and $a=18.6\,\si{\angstrom}$, $b=8.0\,\si{\angstrom}$, $c=7.7\,\si{\angstrom}$, $\alpha=90^o$, $\beta=70^o$, $\gamma=90^o$, Pa symmetry for p(gT2). Dotted lines indicate the peak matches.  }
\end{figure}

\vspace{0.4cm}
When comparing the alkoxylated with the glycolated crystals, we see many differences in packing behaviour. Firstly, changing the side chain changes all parameters of the unit cell for both the monomer and polymer crystals.  Secondly, the $\pi$-stacking characteristics are markedly different.  In the case of p(aT2) the backbones adopt the `tilted stack' while p(gT2) adopts a `deflected stack'. The (0\,1\,0) planes would indicate $\pi$-stacking separations of $5.2$~\si{\angstrom}~and $4.0$~\si{\angstrom}~for p(aT2) and p(gT2) respectively.  Due to thermal disorder and crystal features such as a relatively acute $\alpha$ angle in the p(aT2) crystal, or the non-cofaciality of neighbouring backbones in the p(gT2) crystal, the minimum distance between neighbouring backbones can be as low as $3.4$~\si{\angstrom}~and $3.3$~\si{\angstrom}~respectively. In biological systems, this has been shown to be an important variable in modelling charge transfer rates \cite{winkler1999electron, page1999natural}. Using this measure, as with the $\pi$-stacking distances,  p(gT2) backbones are shown to pack more closely together. 

\vspace{0.4cm}
Interesting features are found in the experimental XRD patterns which can be interpreted with knowledge of the packing motifs. In the case of p(aT2), the surprisingly high intensity of the (3\,0\,0) plane seen in experiment can be attributed to the presence of a strongly diffracting (-1\,0\,1) plane at the same $\vec{q}$ value as the (3\,0\,0) plane. Whilst the relative intensity of this peak is low in the simulated pattern from MD, the difference is likely arising due to the patterns from MD being calculated using the Debye Equation, and therefore does not account for interference terms.    In the case of p(gT2), the strong peak at \\ $\vec{q} = 0.85$ \si{\angstrom}$^{-1}$ can be seen to arise from the (0\,1\,0) plane in the direction of the $\pi$-stack, occurring at double the $\pi$-stack distance due to the Pa symmetry of the cell, as well as arising from the (0\,0\,1) backbone repeat planes. In the case of p(aT2), higher than expected scattering is seen at $\vec{q} = 1.21\, \si{\angstrom}^{-1}$, indicating scattering from other planes as well as the (5\,0\,0) plane.  In the simulations we see a clustering of peaks in this region, any of which may add to intensity in the experimental pattern.  Indices for these peaks are difficult to allocate due to the large number of peaks seen in this region in the p(aT2) structure factor, however for the p(aT2) unit cell we would expect the (0\,1\,0) $\pi$-stack peak to be at $\vec{q} = 1.21 \, \si{\angstrom}^{-1}$, so it is possible it is present giving extra intensity in this region around the (5\,0\,0) plane.  At q-values beyond $\vec{q} = 1.21 \, \si{\angstrom}^{-1}$, peak allocation becomes difficult due to the high number of overlapping peaks in the structure factors and the possibility that side chains have sufficient order to scatter in this region, as discussed previously.

\vspace{0.4cm}
The chain geometries provide a means to estimate the strength of electronic coupling between chains, which will strongly influence the rate of charge transport within crystalline regions of polymer films.  Since both crystal structures have at least C2 symmetry, $\pi$-stack transfer integrals can be approximated via the energy-splitting-in-dimer method \cite{coropceanu2007charge, koopmans1934zuordnung}.  The p(aT2) transfer integrals are found to be 0.011~eV,  and 0.059~eV for HOMO-HOMO and LUMO-LUMO couplings respectively when calculated for two dimers using B3LYP functional and 6-311(g,p) basis set. For p(gT2), the same transfer integrals are found to be 0.058~eV and 0.141~eV.

\vspace{0.4cm}
To elucidate the effect of water exposure on the crystal structure, we immerse both alkoxylated and glycolated crystallites in water, the results of which are seen in {\bf Figure 3}.   In the case of the glycolated oligomers,  water disrupts the crystalline packing, and in the case of gT2, the crystallite has begun to dissolve by the end of the simulation.  For p(gT2) exposure to water causes the side chains to become less ordered, and their end to end lengths shorten and generally adopt a curled structure, causing a reduction in the lamellar stack spacing. Alongside this, upon wetting of p(gT2), the `deflected stack' becomes more deflected, but overall the $\pi$-stack maintains structural order with a larger $\pi$-stack spacing of 6.0~nm (see Figure S19).  For aT2 we see the monomers on the edge of the crystallite rearrange themselves so as to form a capsule, and reduce the area of the energetically unfavourable aT2-water interface.  In the centre of the capsule the monomers are still arranged in a crystalline fashion,  maintaining their side chain interdigitation. In the case of p(aT2), the crystallite is undisturbed after wetting, with order being maintained in both the side chain and in the $\pi$-stack, and structural rearrangements of side chains only occurring on the surface of the crystallite.  The relative effect of water on the crystals is quantitatively understood from the solvation free energies (Figure 3C), where hydrophilic  species typically have three times or more the solvation free energy of their hydrophobic counter parts. 

\begin{figure}[t]
\center
  \includegraphics[width=18cm]{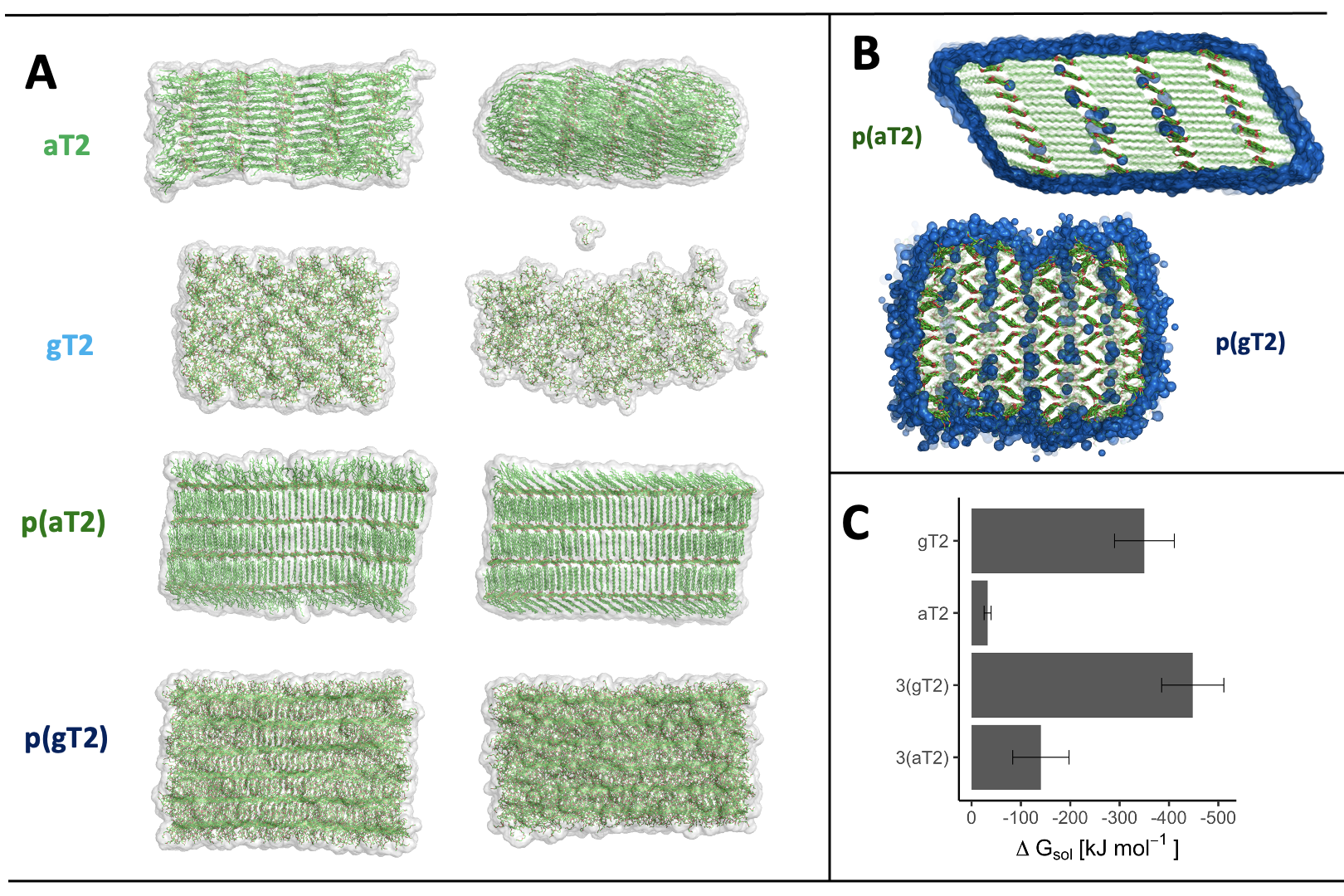}
  \caption{  Structural response of crystallites upon exposure to water. A) Illustrations of the crystallites as viewed on the x-y plane after initial equilibration in MD (left column) and after 50\,ns of simulation time (right column).  The transparent surface indicates the solvent accessible surface areas of the crystallites.  B) Cross sections through the polymer crystallites, showing where and how the water enters, indicated by the blue surface.  Side chains made partially transparent for clarity.  C) Free energy of solvation of alkoxylated and glycolated oligomers presented in this study, calculated using MD (see SI Section 4.2.3). }
\end{figure}

\vspace{0.4cm}
Water is seen to penetrate into both polymer crystallites, with about 4 times more water entering the glycolated crystallite in the 50~ns simulation, as seen in Figure 3B.  In the case of p(gT2) the water enters the lamellar stack, residing amongst the hydrophilic side chains but not causing an increase in the lamellar spacing on this time scale.  Upon inspection it is seen that the water molecules typically sit with their hydrogens coordinating with glycol oxygen atoms, indicating hydrogen bonding occurring in the lamellar stack. Similar behaviour has been observed for poly(ethylene oxide) solutions \cite{smith2000molecular} and in biological environments \cite{brady1993role, lopez2004hydrogen, juranic1996protein}.  For p(aT2), the water resides close to the backbone, furthest from the hydrophobic side chains.  Simulations of bulk wet crystals (see Figure S20) corroborate this observation, showing water droplet formation around the backbone for p(aT2), with each droplet containing around 30 water molecules. These simulations also show that water in crystalline p(gT2) is able to diffuse four times more quickly than in crystalline p(aT2), and both these diffusion speeds are two orders of magnitude smaller than the self-diffusion of water \cite{krynicki1978pressure}.  

\vspace{0.4cm}
It has been observed experimentally that upon charging of a hole-transporting OMIEC,  mass uptake is preceded by mass expulsion,  implying the presence of cations in the bulk before doping \cite{Flagg2020, zhou1987incorporation}. Therefore the impact of ions on the structure of conjugated polymers should be understood.   In order to test the ability of our simulations to correctly reproduce polymer-ion interactions, we first attempted to model the complexes formed between sodium or potassium cations and crown-ether bearing bithiophenes, for which solved crystal structures had previously been obtained. We used this as a test of our force field to correctly model cation chelation in similar materials (Figure S10) \cite{Giovannitti2016b}. In all cases,  our simulations are able to reproduce the experimentally determined crystal structures.   We therefore proceeded to simulate p(aT2) and p(gT2) crystallites immersed in aqueous $0.2\,${\sc m} NaCl. We observed that cations, in the presence of glycol side chains, are becoming localised in space for periods of time, leading to small areas of high cation density ({\bf Figure 4}A).  On inspection we see that this is due to cation-side chain chelation, such as those found for the crown-ether bithiophenes \cite{Giovannitti2016b}.  No such trapping effect is seen for anions,  or for cations with the p(aT2) crystal.

\vspace{0.4cm}
To deepen our understanding of how cations could insert themselves into a crystalline lattice, we quantitatively study cation-glycol chelation using Metadynamics \cite{bonomi2019promoting, tribello2014plumed, bonomi2009plumed}, as shown in Figure 4B.  A free energy landscape is calculated as a function of the distance of the ion from a side chain on one side of the oligomer, $R_1$, and the distance of the ion from a side chain on the other side of the oligomer, $R_2$ (see SI Section 4.2.6 for more details on Metadynamics simulations). In the free energy landscape four notable states are identified.  The first being the solvated state at high $R_1$ and $R_2$.  Two shallow states are observed at $R_1 = 0.1$~nm or $R_2 = 0.1$~nm, corresponding to single bound metastable states to a single side chain with binding energy of  $-0.01 \pm 0.07 \, \text{kJ mol}^{-1}$. Finally a stable state is seen at $R_1 = 0.1 \text{ nm} \, \& \, R_2 = 0.1 \text{ nm}$, corresponding to the ion being double bound to both side chains with energy of $-13.3 \, \pm 0.2 \, \text{kJ mol}^{-1}$.

\begin{figure}[t]
\center
  \includegraphics[width=18cm]{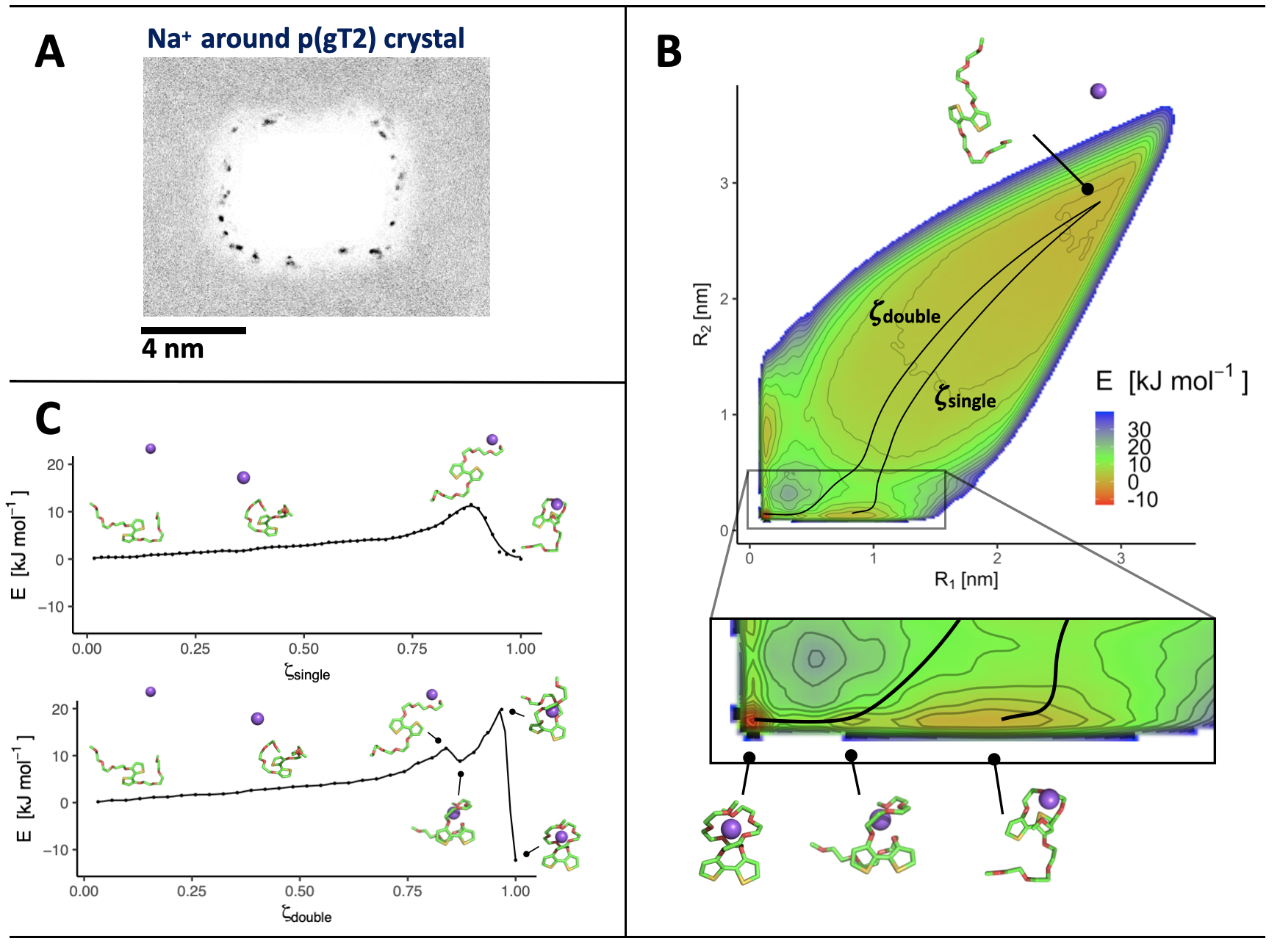}
    \caption{ A) Density plots over the 50 ns production run of Na ions in the p(gT2) crystallite simulation. B) The Free Energy Surface of cation as a function of its distance from a glycol side chain on one side of the polymer ($R_1$), and its distance to a side chain on the other side of the polymer ($R_2$), with the solvated state shown.  Black lines show the minimum energy pathways from the solvated state to the single bound state ($\zeta_\text{single}$) and double bound state ($\zeta_\text{double}$).  Inset shows an area  of the FES containing the single bound state, double bound state and a meta-stable single twisted state.  C) The minimum energy pathway from the solvated state to the single bound state and the double bound state, with illustrations of the bound and transition state configurations.    }
\end{figure}

\vspace{0.4cm}
Shown in Figure 4C are minimum energy transition pathways from the solvated state to the bound states along reaction coordinates $\zeta_\text{single}$ and $\zeta_\text{double}$. We see that for the formation of the double bound state, a larger energy barrier must be overcome compared to the single bound state ($18.5 \pm 0.4 \, \text{kJ mol}^{-1} $ compared to $11.5 \pm 0.4 \,\text{kJ mol}^{-1}$), which can be attributed to the requirement for the backbone to twist and the entropic penalty associated with fitting a second side chain around an ion.  We also simulated the interactions between glycol side chains and the anion, and no notable states were seen (Figure S27).  Finally,  a strong dependence of binding energy is seen with side chain length (Figure S21), in particular between tri-ethylene and tetra-ethylene glycol whereby the single bound state goes from meta-stable to stable, and the double bound state goes from stable to unstable.

\vspace{0.4cm}
Whilst we do not observe ions entering the crystallites on the sub 100~ns time scale, metadynamics has revealed potential ionic structure and dynamics.  In the case of polythiophenes with tail-to-tail side chain attachment, like those studied here,  the double bound state requires the twisting of the backbone.   This makes them unlikely to form in the crystalline phase due to the stabilising $\pi-\pi$ interactions making the backbone twist energy barrier prohibitively large.  As the single bound states are only meta-stable and have an energy barrier within the range of thermal energies, ion-glycol chelations would represent no significant ion transport trap in the crystalline phase for tri-ethylene glycol.  However, in the case of tetra-ethylene glycol,  the single bound state is much deeper (see Figure S21), and so cation trapping in the crystalline phase of tetra-ethylene glycol polymers is possible.  Whilst we would expect both anions and cations to be in the polymer bulk after water exposure,  cations would experience trapping, the strength of which would depend on the side chain length as demonstrated here.

\vspace{0.4cm}
Comparison of the structural response of gT2 and p(gT2) upon wetting has revealed the importance of the $\pi-\pi$ interactions in stabilising the crystal structure of hydrophilic OMIECs.  Whilst the driving forces for formation of the p(gT2) crystallite are both $\pi-\pi$ interactions and electrostatics,  for gT2 it is only the latter due to the absence of $\pi$-stacking.  The dipole interactions are rapidly screened when water is present, thus leading to the rapid dissolution of the gT2 crystallite, and likely also leading to a dependance of molecular weight on stability in water.   Once the water has entered the lamellar stack for p(gT2),  it is surprising that the backbones retain their long range order, whilst the side chains adopt a relatively amorphous structure with the intercalated water.  Such structure could be described as a crystalline gel, where the crystalline backbones form the interconnecting element, and side chains and water form the gel phase \cite{osada2004polymer}.  The move to an amorphous morphology of the side chains causes a reduction in the lamellar stack, and an increase in the $\pi$-stack distance.  For the hydrophobic crystallite we see some diffusion of water into the crystal.  This is possibly enhanced by the use of alkoxy side chains rather than alkyl.  Furthermore,  if the water enters the p(aT2) crystallite along the backbone, we might expect water entry to be suppressed if the crystallite is surrounded by amorphous bulk polymer, as these narrow channels would become fully or partially blocked.

\vspace{0.4cm}
 Whilst in the solid state some hydrophobic conjugated polymers with very short side chains have been shown to adopt `tilted stacks' \cite{arosio2007first, arosio2009ordered}, the majority have shown to adopt `straight stacks' \cite{tashiro1997crystal, Winokur1991, hugger2004semicrystalline, brinkmann2006orientation, wu2010temperature, dudenko2012strategy}. It is interesting therefore,  that the exchange from an alkyl to an alkoxy side chain induces the `tilted stack'.  This is possibly due to the electrostatic repulsion of the oxygen atoms in the alkoxy side chain, as well as the increased rotational freedom around the base of the side chain associated with exchanging $\text{CH}_2$ groups for oxygen \cite{Meng2015}.  In literature it has been shown that linear alkyl side chains tend to pack with straight interdigitating configurations. Transitioning to a glycol side chain causes the backbones to adopt a `deflected stack'. It is important to note that the glycol crystal more closely resembles an ionic lattice,  so as to place negatively charged atoms opposite positively charged ones. The added strain applied to the system to satisfy this requirement is likely influencing the resulting backbone stacking motif.  Whilst in the monomer crystal the hydrophilic side chains adopt a curled structure,  in the polymer they adopt a gentle s-bend configuration.  The difference in side chain and backbone structure between the monomer and polymer crystals is likely linked to the difference in effective side chain attachment density and relative polarity of their environment, which are both lower for the monomer due to the bulky hydrophobic $\text{Sn(Me)}_3$ end groups.  

\vspace{0.4cm}
The change in backbone packing when exchanging alkoxy for glycol side chains leads to a higher inter-backbone electronic transfer integral.  Given that charge transport is likely dominated by inter-molecular charge transfer \cite{himmelberger2015engineering},  we would propose that, if the impact of side chain type on electrostatic interactions can be neglected,  a perfect dry glycolated crystal is better suited to electronic charge transport than an alkoxylated one. However, our simulations do not sample the crystalline imperfections shown by the broad peaks in the XRD pattern, and likely vary from the alkoxy to glycol polymer materials.

\vspace{0.4cm}
Glycol side chains have been used extensively to improve OMIEC device performance, despite issues remaining around device stabilities during electrochemical cycling. Already design strategies have been proposed to negate these side effects, and usually revolve around altering side chain lengths \cite{moser2020ethylene, Moser2020}, attachment positions \cite{hallani2021regiochemistry}, or combining hydrophobic and hydrophilic moieties \cite{szumska2021reversible, Giovannitti2018, liu2021amphipathic, maria2021effect}.  In all these cases the amorphous phase, as well as the crystalline phase, is likely to be important in maintaining structural integrity of films. Our force field opens up the opportunity to atomistically characterise structural packing and probe the short range order and interactions in the amorphous phase when these chemical designs are applied.  

\vspace{0.4cm}
In conclusion, we have atomistically characterised the polymer packing of archetypal alkoxy and glycolated polythiophene polymers. We uncovered the changes in packing behaviours to accommodate the side chains of different chemistries, in particular the adoption of either a `straight stack', `tilted stack' or a more densely packed `deflected stack' depending on the oxygen content of the side chains. These behaviours indicate the glycolated polymers pack more effectively for optimising charge transport in the solid state. We have subsequently observed the response of the polymer to water exposure, as well as the placement of water in the polymer lattices - around the backbone for alkoxy systems, and in the lamellar stack for glycol systems.  We have shown the important effects of $\pi-\pi$ interactions in stabilising the polymer lattice, demonstrating that enhancing the $\pi$-stacking in systems is critical to ensuring their stability in aqueous environments, with implications for optimising polymer molecular weights.  Finally, metadynamics has revealed that tri-ethylene glycol side chains are the optimum length for ensuring no cation trapping in the crystalline phase of OMIECs,  whilst tetra-ethylene glycol is more suited to applications where cation trapping is desirable.  Atomistic and quantitative analysis, like what we show here, is critical to effectively guiding future polymer OMIEC design. 

\small{
\vspace{0.4cm}
\subsection*{Acknowledgements}
J.N., N. S., D. P. and A. G. acknowledge funding from the European Research Council (ERC) under the European Union's Horizon 2020 research and innovation program (grant agreement no 742708, project CAPaCITy). \\
\vspace{0.2cm}
H. Y. acknowledges support of the Chinese Scholarship Council through a PhD studentship. \\
\vspace{0.2cm}
S. M. T. acknowledges support from the UK Engineering and Physical Sciences Research Council via the Global Challenges Research Fund through the “SUNRISE” project. \\
\vspace{0.2cm}
J. N. thanks the Royal Society for award of a Research Professorship. \\
\vspace{0.2cm}
J. M. F. is supported by a Royal Society University Research Fellowship (URF-R1-191292). \\
\vspace{0.2cm}
C. C. gratefully acknowledges financial support from the National Science Foundation DMR Award (no 1808401) \\
\vspace{0.2cm}
A. G. acknowledges funding from the TomKat Center for Sustainable Energy at Stanford University \\
\vspace{0.2cm}
G. L.  acknowledges support from the National Science Foundation Graduate Research Fellowship Program under grant DGE-1656518 \\
\vspace{0.2cm}
R. S. , R. K. H. and I. M. acknowledge financial support from KAUST, Office of Sponsored Research (OSR) awards no. OSR-2019-CRG8-4086 and OSR-2018-CRG7-3749 \\
\vspace{0.2cm}
R. S. , R. K. H. and I. M. acknowledge funding from the European Union's Horizon 2020 research and innovation program under grant agreement no. 952911, project BOOSTER and grant agreement no. 862474, project RoLAFLEX, EPSRC Project EP/T026219/1 as well as EPSRC Project EP/T004908/1
}

\vspace{0.8cm}
\begin{figure}[h!]
\textbf{Table of Contents}\\
\medskip
  \includegraphics[width=11cm]{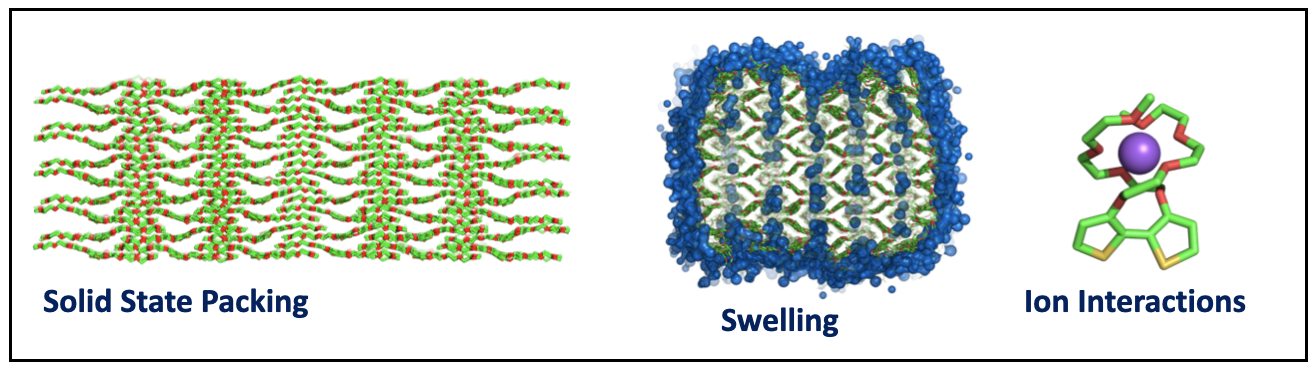}
  \medskip
  \caption*{ We use Molecular Dynamics and X-ray Diffraction to characterise mixed-transport conjugated polymers packing, ability to swell and chelate with ions. We show that glycol side chains can induce tilted $\pi$-stacks, allow water flow into the lamellar stack, and form single and double bound chelates. }
\end{figure}

\clearpage
\bibliographystyle{MSP}
\bibliography{Bib_File}

\end{document}



\title{Impact of Side Chain Hydrophilicity on Packing, Swelling and Ion Interactions in Oxy-bithiophene Semiconductors - Supporting Information}
\date{}
\maketitle

\clearpage
\tableofcontents



\newpage
\section{Packing Motifs}

\begin{figure}[h!]
\center
  \includegraphics[width=15cm]{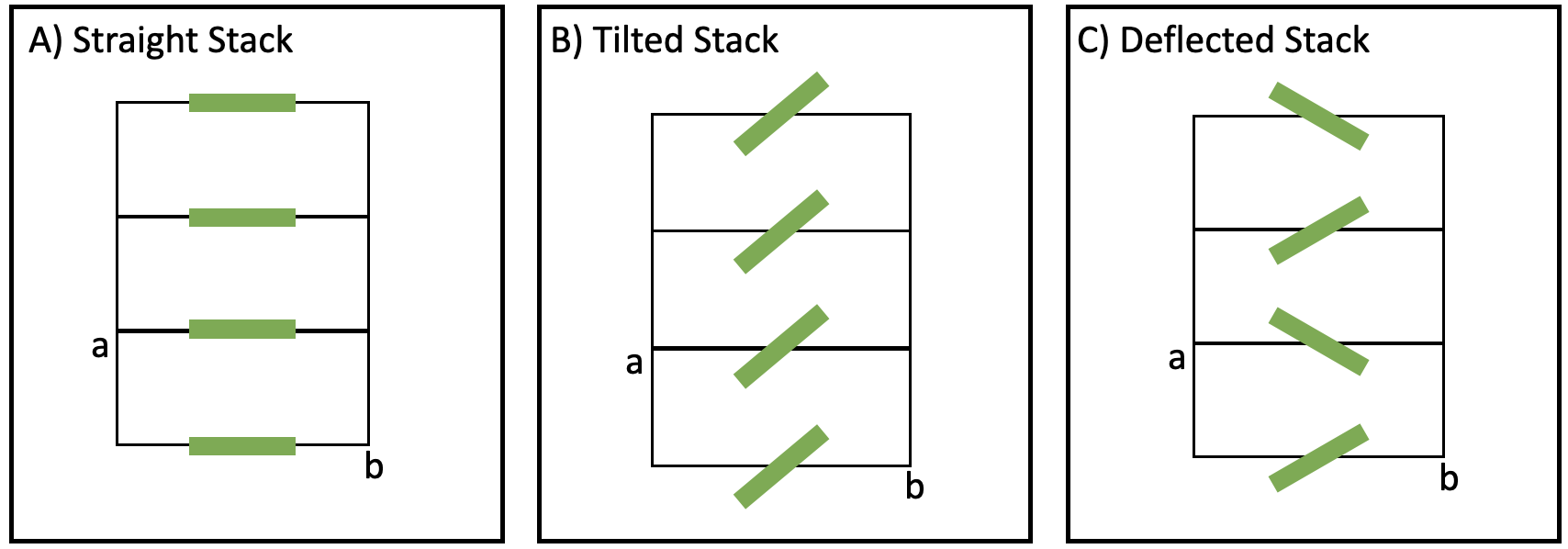}
 \caption{ Backbone packing motifs referred to in this study. Illustrations with a view along the polymer backbone.  Terminology is that adopted by K. Tashiro \textit{et. al.} \cite{tashiro1997crystal}. }
   \label{fig:Chem_Formula}
\end{figure}

\section{Molecular Dynamics Parameters}

Presented here are atomistic MD parameters for the monomers and polymers simulated in this study. 

Our MD force field is based on OPLS-AA \cite{Jorgensen1996}, with inter-thiophene dihedrals parameterised to fit with density functional theory (DFT) at the B3LYP/6311G** level of theory (Figure S2 and S3) \cite{Wildman2016}.  Glycol side chain parameters are from work by D. Pearce \cite{woods2020side} and are based on OPLS-aa atom types, with other backbone bonded parameters from work by M. Moreno {\it et. al.} and R. Bhatta {\it et. al.} \cite{Moreno2010, Bhatta2013}. 

 Ions parameters are taken from OPLS, and are not shown here explicitly.  Water is parameterised according to the OPC model \cite{Izadi2014}. Small molecules used in the simulation of ion crystals are from OPLS, except for partial charges, $q$, which are fitted to B3LYP/6311g** using CHELPG mapping, and are shown in the molecular topologies section. 

\clearpage
\subsection{Atom Types}

Atom types are based on OPLS.  The type determines the non-bonded Lennard-Jones (LJ) interactions and the bonded potentials. LJ interactions between atoms $i$ and $j$ follow 
\begin{equation}
V_{LJ}(\vec{r}_{ij}) = 4 \, \epsilon_{ij} \, \Big(  \Big( \frac{\sigma_{ij}}{r_{ij}} \Big)^{12} -  \Big( \frac{\sigma_{ij}}{r_{ij}} \Big)^{6} \Big) \, .
\end{equation}
To find the interaction parameters $\sigma_{ij}$ and $\epsilon_{ij}$, we follow the standard set by the OPLS force field and use the mixing rules $\sigma_{ij} = (\sigma_{ii} \, \sigma_{jj})^{1/2}$ and $\epsilon_{ij} = (\epsilon_{ii} \, \epsilon_{jj})^{1/2}$.  Atom types and LJ parameters for ions and small molecules used in simulation of ion crystals are found in the directly in the OPLS forcefield. 

\vspace{0.5cm}
\begin{center}
	\scalebox{0.9}{
    \begin{tabular}{  p{2.4cm} p{2.5cm} p{2.5cm} p{2.5cm} }
Type & Mass [a.u] &  $\sigma_{ii}$ [nm] & $\epsilon_{ii}$ $[\text{kJ mol}^{-1}]$ \\ \specialrule{1.5pt}{1pt}{1pt}
 \multicolumn{2}{l}{Thiophenes}   &  &    \\ \hline
CAA   & 12.0110 & 0.355 & 0.293   \\ 
CE    & 12.0110 & 0.355 & 0.293   \\ 
CBB   & 12.0110 & 0.355 & 0.293   \\ 
HT    &  1.0080 & 0.242 & 0.126   \\ 
ST    & 32.0600 & 0.355 & 1.046   \\ 
OT    & 15.9994 & 0.307 & 0.711 \\ \hline
 \multicolumn{2}{l}{End Groups}   &  &    \\ \hline
BT    & 79.9040 & 0.347 & 1.977   \\ 
SnT   & 118.710 & 0.367 & 2.420  \\ 
CSn   & 12.0110 & 0.350 & 0.276   \\ 
HSn   &  1.0080 & 0.250 & 0.126  \\ \hline 
 \multicolumn{2}{l}{Glycol Side Chains}   &  &    \\ \hline
CG!  & 12.0110 & 0.350 & 0.276   \\ 
CG   & 12.0110 & 0.350 & 0.276   \\ 
HG   &  1.0080 & 0.250 & 0.126   \\ 
OG   & 15.9994 & 0.290 & 0.586   \\ 
CL!  & 12.0110 & 0.350 & 0.347  \\ 
CL   & 12.0110 & 0.350 & 0.347  \\ 
HL   & 1.00800 & 0.250 & 0.126  \\  
    \end{tabular}\label{scatLen}}
    \captionof{table}{ Atom types } 
\end{center}

\newpage
\subsection{Bonded Parameters}
\subsubsection{Bonds}

Bonds are modelled using a harmonic potential,
\begin{equation}
V_{bond}(\vec{r}_{ij}) = \frac{1}{2} \, k^b_{ij} \, (r_{ij} - b^0_{ij})^2 .
\end{equation}

\vspace{1cm}
\begin{center}
\scalebox{0.8}{
    \begin{tabular}{  p{2.6cm} p{2.6cm} p{2.6cm} p{3.6cm} }
 $i$ Type &  $j$ Type &  $b^0_{ij}$ [nm] & $k^b_{ij}$ $[\text{kJ mol}^{-1}\text{nm}^{-2}]$ \\ \specialrule{1.5pt}{1pt}{1pt}
 \multicolumn{2}{l}{Thiophenes}   &  &    \\ \hline
 ST  & CAA & 0.1760  & 209200.0    \\ 
 ST  & CE  & 0.1760  & 209200.0    \\ 
 CAA & CBB & 0.1400  & 392459.2    \\ 
 CE  & CBB & 0.1400  & 392459.2    \\ 
 CBB & CBB & 0.1400  & 392459.2    \\ 
 CBB & HT  & 0.1080  & 307105.6    \\ 
 CAA & HT  & 0.1080  & 307105.6    \\ 
 CE  & HT  & 0.1080  & 307105.6    \\ 
 CAA & CAA & 0.1400  & 392459.2    \\ 
 CE  & CE  & 0.1400  & 392459.2    \\ 
 CBB & OT  & 0.1364  & 376560.0    \\ \hline 
 \multicolumn{2}{l}{End Groups}   &  &    \\ \hline
 CE  & BT   & 0.1870  & 251040.0    \\ 
 CE  & SnT  & 0.2130  & 230120.0    \\ 
 CSn & SnT  & 0.2142  & 230120.0    \\ 
 CSn & HSn  & 0.1090  & 284512.0    \\ \hline
 \multicolumn{2}{l}{Glycol Side Chains}   &  &    \\ \hline
 OT   & CG! & 0.1410 & 267776.0 \\
 CG!  & CG & 0.1529 & 224262.4 \\ 
 CG!  & HG & 0.1090 & 284512.0 \\
 CG   & CG & 0.1529 & 224262.4 \\ 
 CG   & HG & 0.1090 & 284512.0 \\
 CG   & OG & 0.1410 & 267776.0 \\  \hline
 \multicolumn{2}{l}{Alkyl Side Chains}   &  &    \\ \hline
 OT  & CL! & 0.1410 & 267776.0 \\ 
 CL!  & CL & 0.1529 & 224262.4 \\
 CL   & CL & 0.1529 & 224262.4 \\
 CL!  & HL & 0.1090 & 284512.0 \\
 CL   & HL & 0.1090 & 284512.0 \\

    \end{tabular}\label{scatLen}}
    \captionof{table}{ Bond Parameters } \vspace{0.3cm}
\end{center}

\newpage
\subsubsection{Angles}

Angles are modelled using a harmonic potential, 
\begin{equation}
V_{angle}(\theta_{ijk}) = \frac{1}{2} \, k^\theta_{ijk} \, (\theta_{ijk} - \theta^0_{ijk})^2 .
\end{equation}
Angle parameters for thiophene rings are from Moreno {\it et. al.} \cite{Moreno2010}. Angle parameters for side chains are from work by Drew Pearce \cite{woods2020side}. Due to the oxygen in the two position on the thiophene rings likely forming part of the conjugated network,  OT-CBB-CBB and OT-CBB-CAA angles are parameterised to fit calculations from density functional theory at the B3LYP/6311g** level of theory, see Figure S2. 

\begin{figure}[h!]
\center
  \includegraphics[width=7cm]{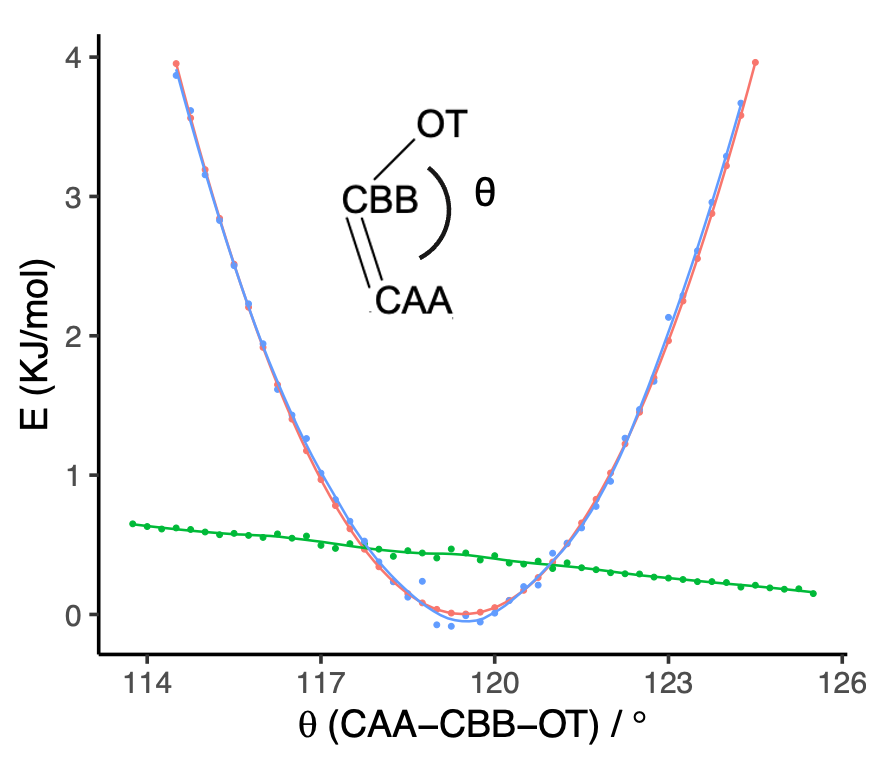}
 \caption{ Angle parameterisation for CAA-CBB-OT angle.  Green line shows the potential in MD with the angle DOF switched off.  The Red line indicates the result from DFT, and the blue line is the MD potential with the angle DOF turned on after fitting to the difference in the red and green lines.  }
   \label{fig:Chem_Formula}
\end{figure}

\vspace{1cm}
\begin{center}
\scalebox{0.8}{
    \begin{tabular}{  p{2.3cm} p{2.3cm} p{2.3cm} p{2.3cm} p{3.6cm} }
 $i$ Type &  $j$ Type &  $k$ Type &  $\theta^0_{ijk}$ [$^o$] & $k^\theta_{ijk}$ $[\text{kJ mol}^{-1} \, ^{o-2}]$ \\ \specialrule{1.5pt}{1pt}{1pt}
 \multicolumn{2}{l}{Thiophenes}   &  &  &   \\ \hline
 CE   & ST  & CAA  &  91.0 & 502.08  \\ 
 ST   & CAA & CBB  & 112.0 & 502.08  \\ 
 ST   & CE  & CBB  & 112.0 & 502.08  \\ 
 CAA  & CBB & CBB  & 112.5 & 502.08  \\ 
 CE   & CBB & CBB  & 112.5 & 502.08  \\ 
 CBB  & CBB & HT   & 124.0 & 292.88  \\ 
 CBB  & CE  & HT   & 123.5 & 292.88  \\ 
 ST   & CE  & HT   & 112.0 & 502.08  \\ 
 CAA  & CAA & CBB  & 129.0 & 502.08  \\ 
 CE   & CE  & CBB  & 129.0 & 502.08  \\ 
 HT   & CBB & CE   & 123.5 & 502.08  \\ 
 ST   & CAA & CAA  & 119.0 & 502.08  \\ 
 ST   & CE  & CE   & 119.0 & 502.08  \\ 
 OT   & CBB & CBB  & 127.5 & 870.00  \\ 
 CAA  & CBB & OT   & 119.0 & 850.00  \\ \hline
  \multicolumn{2}{l}{End Groups}   &  &  &   \\ \hline
 CBB  & CE  & BT  & 117.42 & 627.600 \\
 ST   & CE  & BT  & 130.19 & 627.600 \\
 CBB  & CE  & SnT & 121.84 & 627.600 \\
 ST   & CE  & SnT & 125.73 & 627.600 \\
 CSn  & SnT & CSn &  97.86 & 518.816 \\
 CE   & SnT & CSn & 111.40 & 518.816 \\
 HSn  & CSn & HSn & 109.49 & 276.144 \\
 SnT  & CSn & HSn & 109.43 & 292.880 \\ \hline
  \multicolumn{2}{l}{Glycol Side Chains}   &  &  &   \\ \hline
OT  & CG! & CG	& 109.500 & 418.400 \\
CG  & CG  & OG	& 109.500 & 418.400 \\
CG  & OG  & CG	& 109.500 & 502.080 \\
CG  & CG  & HG	& 110.700 & 313.800 \\
OG  & CG  & HG	& 109.500 & 292.880 \\
HG  & CG  & HG	& 107.800 & 276.144 \\
HG  & CG! & HG	& 107.800 & 276.144 \\
HG  & CG! & CG	& 110.700 & 313.800 \\
CG! & CG  & HG	& 110.700 & 313.800 \\
CG! & CG  & OG	& 109.500 & 418.400 \\ \hline
  \multicolumn{2}{l}{Alkyl Side Chains}   &  &  &   \\ \hline
CL  & CL  & CL	 & 112.700 & 488.273 \\
CL  & CL  & HL	 & 110.700 & 313.800 \\
CL  & CL! & HL	 & 110.700 & 313.800 \\
HL  & CL  & HL	 & 107.800 & 276.144 \\
CL! & CL  & CL	 & 112.700 & 488.273 \\
CL! & CL  & HL	 & 110.700 & 313.800 \\
HL  & CL! & HL	 & 107.800 & 276.144 \\
OT  & CL! & CL	 & 109.500 & 418.400 \\
       \end{tabular}\label{scatLen}}
    \captionof{table}{ Angle Parameters } \vspace{0.3cm}
\end{center}

\newpage
\subsubsection{Improper Dihedrals}

Improper dihedrals are employed here to keep conjugated groups planar, and are modelled with a harmonic potential,
\begin{equation}
V_{Improper}(\zeta_{ijkl}) = \frac{1}{2} \, k^\zeta_{ijkl} \, (\zeta_{ijkl} - \zeta^0_{ijkl})^2 .
\end{equation}
In this work we set $\zeta^0_{ijkl} = 0$ and $k^\zeta_{ijkl} = 391.47 \,  \text{kJ mol}^{-1 \, o \, -2}$.

\subsubsection{Proper Dihedrals}

Proper dihedrals are modelled using Ryckaert-Bellemans functions,
\begin{equation}
V_{Proper}(\phi_{ijkl}) = \sum^5_{n=0} \, C_n (cos(\phi_{ijkl}))^n .
\end{equation}
Proper dihedral parameters are from work by R. Bhatta {\it et. al.} \cite{Bhatta2013} and Drew Pearce \cite{woods2020side} , except for dihedrals around the thiophene rings, which are fitted to DFT, see Figure 3.
\begin{figure}[h!]
\center
  \includegraphics[width=15cm]{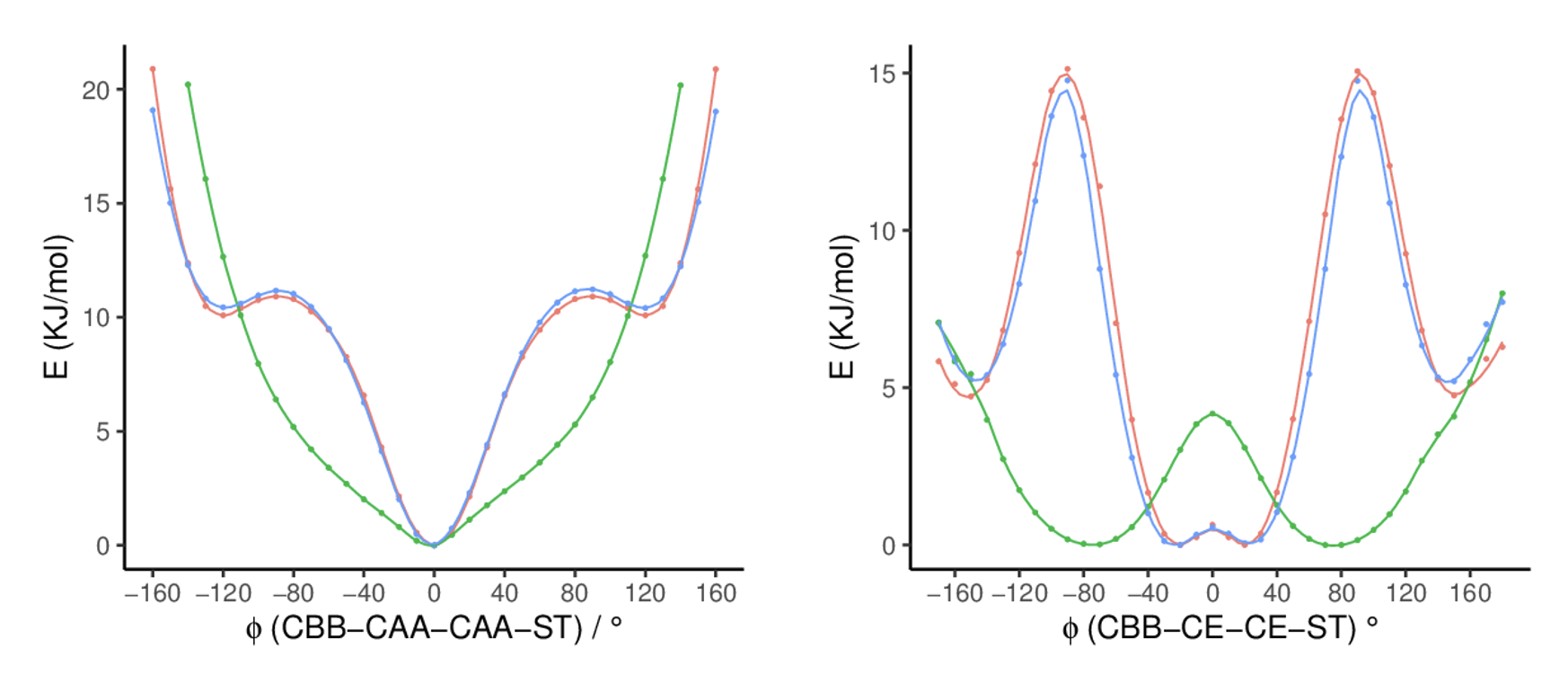}
 \caption{ Dihedral parameterisation for the intra-monomer dihedral (left), and inter-monomer dihedral (right).  Green line shows the potential in MD with the angle DOF switched off.  The Red line indicates the result from DFT, and the blue line is the MD potential with the angle DOF turned on after fitting to the difference in the red and green lines.  }
   \label{fig:Chem_Formula}
\end{figure}

\vspace{1cm}
\begin{center}
\scalebox{0.8}{
    \begin{tabular}{  p{1.4cm} p{1.4cm} p{1.4cm} p{1.4cm} p{1.8cm} p{1.8cm} p{1.8cm} p{1.8cm} p{1.8cm} p{1cm} }
 $i$ Type &  $j$ Type &  $k$ Type &  $l$ Type &  $C_0$ & $C_1$ & $C_2$ & $C_3$ & $C_4$ & $C_5$  \\ \specialrule{1.5pt}{1pt}{1pt}
 \multicolumn{2}{l}{Thiophenes}   &  &  &  & & & & & \\ \hline
 CBB   & CAA  & ST   & CAA  & 528.5228 & -459.4450   & -82.5837  & -105.8677 & 119.3695 & 0.0000 \\ 
 CAA   & ST   & CAA  & CBB  & 528.5228 & -459.4450   & -82.5837  & -105.8677 & 119.3695 & 0.0000 \\ 
 CAA   & CBB  & CBB  & CAA  & 528.5228 & 459.4450    & -82.5837  & -105.8677 & 119.3695 & 0.0000 \\ 
 CBB   & CBB  & CAA  & ST   & 528.5228 & -459.4450   & -82.5837  & -105.8677 & 119.3695 & 0.0000 \\ 
 OT    & CBB  & CBB  & CAA  & 492.2476 & 996.8798   & 861.7366  & 471.9970  & 114.9219 & 0.0000 \\ 
 CBB   & CBB  & CAA  & CAA  & 316.2894 & 485.3440    & 178.5689  & -6.3931   & -15.9565 & 0.0000 \\ 
 CAA   & ST   & CAA  & CAA  & 12.3566  & -0.6573     & -17.7091  & -1.6648   & 7.8889	& 0.0000 \\ \hline 
 \multicolumn{2}{l}{End Groups}   &  &  &  & & & & & \\ \hline
 CBB   & CE   & SnT  & CSn  & 1.3284   & 4.7153	     & 59.1743	 & -93.2906  & 28.1114	& 0.0000 \\ 
 CE    & SnT  & CSn  & HSn  & 1.3284   & 4.7153	     & 59.1743	 & -93.2906  & 28.1114	& 0.0000 \\ \hline 
 \multicolumn{2}{l}{Glycol Side Chain}   &  &  &  & & & & & \\ \hline
  CBB   & CBB  & OT  & CG!  &  1.3284   & 4.7153  & 59.1743 & -93.2906 & 28.1114  & 0.0000 \\  
  CBB   & OT   & CG! & CG   &  1.7150   & 2.8450  & 1.046   & -5.6070  & -0.0000  & 0.0000 \\  
  CBB   & OT   & CG! & HG   &  1.7150   & 2.8450  & 1.046   & -5.6070  & -0.0000  & 0.0000 \\  
  CAA   & CBB  & OT  & CG!  &  1.3284   & 4.7153  & 59.1743 & -93.2906 & 28.1114  & 0.0000 \\  
  OT    & CG!  & CG  & OG   & -1.1510   & 1.1510  & 0.000   & -0.0000  & -0.0000  & 0.0000 \\  
  CG    & CG  & OG  & CG  &  1.715   &	2.845   & 1.046 & -5.607 & -0.000  & 0.0000 \\  
  CG!   & CG  & OG  & CG  &  1.715   &	2.845   & 1.046 & -5.607 & -0.000  & 0.0000 \\  
  HG    & CG  & CG  & HG  &  0.628   &	1.883   & 0.000 & -2.510 & -0.000  & 0.0000 \\  
  HG    & CG  & CG  & OG  &  0.979   &	2.937   & 0.000 & -3.916 & -0.000  & 0.0000 \\  
  HG    & CG  & OG  & CG  &  1.590   &	4.770   & 0.000 & -6.360 & -0.000  & 0.0000 \\  
  OG    & CG  & CG  & OG  & -1.151   &	1.151   & 0.000 & -0.000 & -0.000  & 0.0000 \\  \hline
 \multicolumn{2}{l}{Alkyl Side Chain}   &  &  &  & & & & & \\ \hline
  CBB    & CBB  & OT   & CL!   &  1.3284   & 4.7153  & 59.1743   & -93.2900 & 28.111  & 0.0000 \\  
  CBB    & OT   & CL!  & CL    &  1.7150   & 2.8450  & 1.0460    & -5.6070  & -0.0000 & 0.0000 \\  
  CBB    & OT   & CL!  & HL    &  1.7150   & 2.8450  & 1.0460    & -5.6070  & -0.0000 & 0.0000 \\  
  CAA    & CBB  & OT   & CL!   &  1.3284   & 4.7153  & 59.1743   & -93.2900 & 28.111  & 0.0000 \\  
  OT     & CL!  & CL   & CL    & -1.1510   & 1.1510  & 0.0000    & -0.0000  & -0.0000 & 0.0000 \\  
  CL	    & CL	  & CL  & CL   & 2.9288   & -1.4644  & 0.209   & -1.6736 & 0.0000  & 0.0000 \\  
  CL	    & CL	  & CL  & HL   & 0.6276   & 1.88280  & 0.000   & -2.5104 & 0.0000  & 0.0000 \\  
  HL	    & CL	  & CL  & HL   & 0.6276   & 1.88280  & 0.000   & -2.5104 & 0.0000  & 0.0000 \\  
  CL!	    & CL	  & CL  & CL   & 2.9288   & -1.4644  & 0.209   & -1.6736 & 0.0000  & 0.0000 \\  
  HL	    & CL!	  & CL  & HL   & 0.6276   & 1.88280  & 0.000   & -2.5104 & 0.0000  & 0.0000 \\  
  CL	    & CL!	  & CL  & CL   & 2.9288   & -1.4644  & 0.209   & -1.6736 & 0.0000  & 0.0000 \\  
  OG	    & CG	  & CG  & CL   & -1.151   & 1.151    & 0.000   & -0.000  & 0.0000  & 0.0000 \\

       \end{tabular}\label{scatLen} }
    \captionof{table}{ Proper Dihedral Parameters }  
\end{center}

\newpage
\subsection{Atom Naming Schemes}

\begin{figure}[h!]
\center
  \includegraphics[width=17cm]{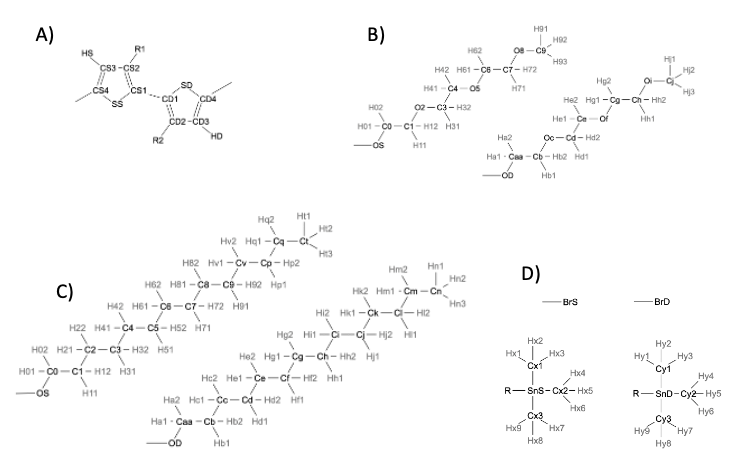}
 \caption{ Atom naming schemes for a) thiophene backbone, b) glycol side chains, c) alkyl side chains, and d) end groups.  }
   \label{fig:Chem_Formula}
\end{figure}

\newpage
\subsection{Atom Type Schemes}

\begin{figure}[h!]
\center
  \includegraphics[width=15cm]{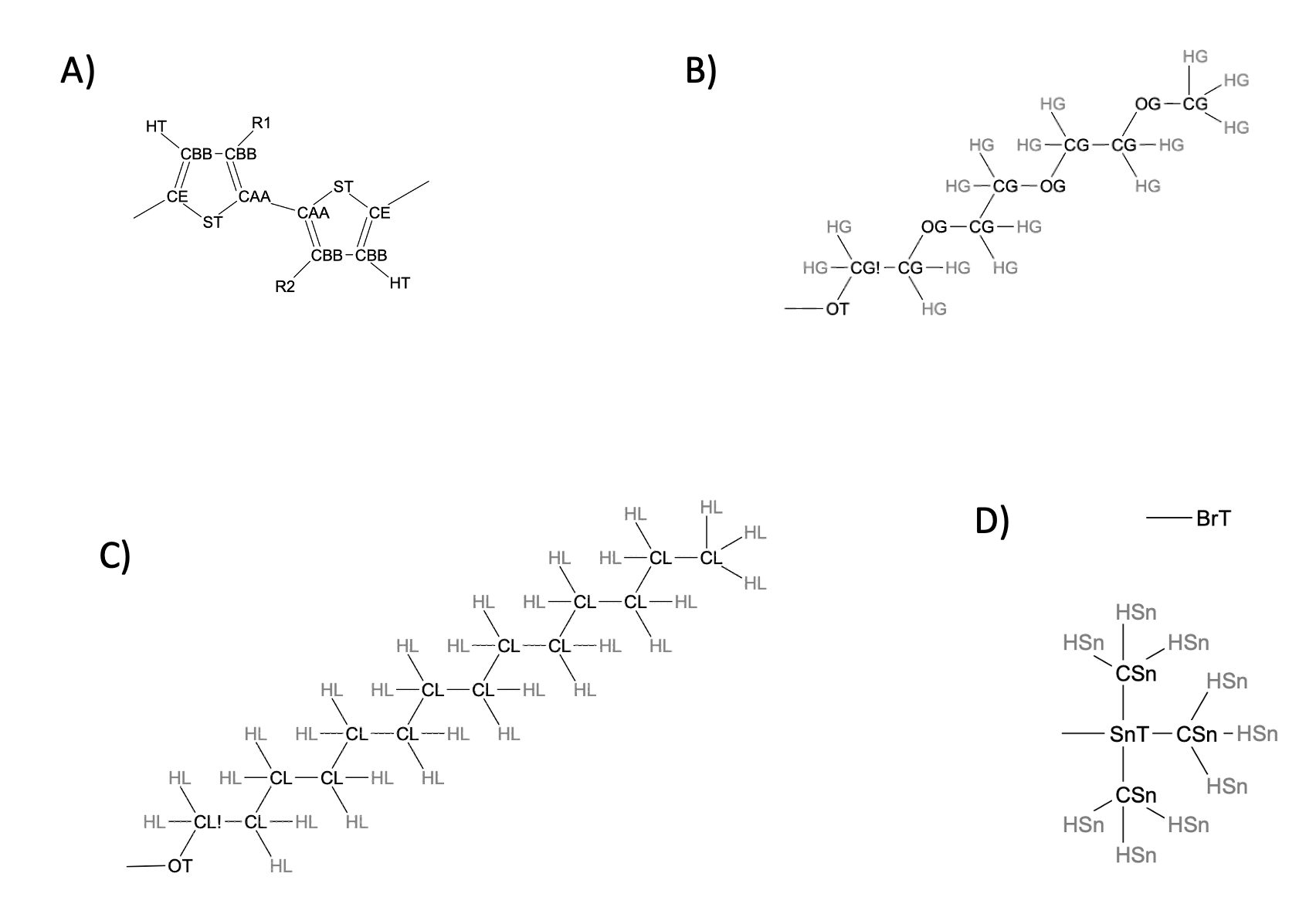}
 \caption{ Atom type schemes for a) thiophene backbone, b) glycol side chains, c) alkyl side chains, and d) end groups.  }
   \label{fig:Chem_Formula}
\end{figure}

\newpage
\subsection{Electrostatics}

Partial charges for monomers are calculated for the entire molecule.  In the case of polymers, partial charges are calculated for 5mers, where the central monomer has side chains, and the rest have their side chains removed.  Backbone partial charges for alkyl or glycol species were very similar, and so were averaged across the two so as to make the forcefield more generalisable.  Glycol side chain partial charges are calculated in a similar way in previous work by Drew Pearce \cite{woods2020side}, and are taken from that work. 

\vspace{1cm}
\begin{table}[h!]
\begin{minipage}{.5\linewidth}
\scalebox{0.8}{
  \begin{tabular}[htbp]{@{}lll@{}}
    Atom Name  & $q_{\text{monomer}}$ & $q_{\text{polymer}}$ \\ \specialrule{1.5pt}{1pt}{1pt}
    Thiophenes &   &   \\
    \hline
    CS1, CD1  & -0.123   & -0.004 \\
    CS2, CD2  & +0.393   & +0.233 \\
    CS3, CD3  & -0.291   & -0.347 \\
    CS4, CD4  & -0.108   & +0.119 \\
    SS, SD      & +0.051   & -0.198 \\
    HS, HD      & +0.170   & +0.221 \\
    \hline
    Alkyl Side Chain  &   &   \\
    \hline
   OS, OD	 &    +0.208     & -0.208 \\
   C0, Caa    	 & +0.271  & +0.070  \\
   H01,H02,Ha1,Ha2   	 & +0.014  & +0.014  \\
   C1, Cb    	 & -0.099  & +0.122  \\
   H11,H12,Hb1,Hb2   	 & +0.033  & -0.008  \\
   C2, Cc    	 & -0.112  & -0.039  \\
   H21,H22,Hc1,Hc2   	 & +0.049  & +0.004  \\
   C3, Cd    	 & -0.034  & +0.026  \\
   H31,H23,Hd1,Hd2   	 & +0.018  & -0.015  \\
   C4, Ce    	 & +0.037  & +0.067  \\
   H41,H42,He1,He2   	 & +0.013  & -0.024  \\
   C5, Cf    	 & -0.149  & +0.024  \\
   H51,H52,Hf1,Hf2   	 & +0.047  & -0.017  \\
   C6, Cg    	 & -0.063  & +0.075  \\
   H61,H62,Hg1,Hg2   	 & +0.017  & -0.031  \\
   C7, Ch    	 & +0.201  & +0.037  \\
   H71,H72,Hh1,Hh2   	 & -0.035  & -0.027  \\
   C8, Ci    	 & -0.117  & +0.040  \\
   H81,H82,Hi1,Hi2   	 & +0.017  & -0.021  \\
   C9, Cj    	 & -0.061  & +0.100  \\
   H91,H92,Hj1,Hj2   	 & +0.006  & -0.035  \\
   Cv, Ck    	 & +0.184  & -0.025  \\
   Hv1,Hv2,Hk1,Hk2   	 & -0.030  & -0.011  \\
   Cp, Cl    	 & -0.131  & +0.040  \\
   Hp1,Hp2,Hl1,Hl2   	 & +0.020  & -0.014  \\
   Cq, Cm    	 & +0.199  & +0.227  \\
   Hq1,Hq2,Hm1,Hm2   	 & -0.036  & -0.058  \\
   Ct, Cn    	 & -0.285  & -0.157  \\
   Ht1,Ht2,Ht3,Hn1,Hn2,Hn3   	 & +0.062  & +0.021  \\
  \hline
  \end{tabular}}
  \end{minipage}%
\begin{minipage}{.5\linewidth}
\scalebox{0.8}{
  \begin{tabular}[htbp]{@{}lll@{}}
    Atom Name  & $q_{\text{monomer}}$ & $q_{\text{polymer}}$ \\ \specialrule{1.5pt}{1pt}{1pt}
  Glycol Side Chain  &   &   \\
  \hline
   OS, OD			 &  -0.563	&  -0.208 \\
   C0, Caa    			 &  +0.196	&  -0.236 \\
   H01,H02,Ha1,Ha2   		 &  +0.049	&  +0.100 \\
   C1, Cb    			 &  +0.515	&  +0.023 \\
   H11,H12,Hb1,Hb2   		 &  -0.018	&  +0.100 \\
   O2, Oc    			 &  -0.797	&  -0.373 \\
   C3, Cd    			 &  +0.357	&  +0.015 \\
   H31,H23,Hd1,Hd2   		 &  -0.015	&  +0.100 \\
   C4, Ce    			 &  +0.575	&  +0.019 \\
   H41,H42,He1,He2   		 &  -0.046	&  +0.100 \\
   O5, Of    			 &  -0.784	&  -0.373 \\
   C6, Cg    			 &  +0.340	&  +0.019 \\
   H61,H62,Hg1,Hg2   		 &  -0.015	&  +0.100 \\
   C7, Ch    			 &  +0.371	&  +0.012 \\
   H71,H72,Hh1,Hh2	   	 &  +0.005	&  +0.100 \\
   O8, Oi    			 &  -0.548	&  -0.373 \\
   C9, Cj    			 &  -0.017	&  -0.039 \\
   H91,H92,H93,Hj1,Hj2Hj3   	 &  +0.077	&  +0.100 \\ \hline 
    End Groups &   &   \\
    \hline
    SnS, SnD  & -1.141   & - \\
    Cx1, Cy1 &  -0.784  & - \\
    Hx1, Hx2, Hx3, Hy1, Hy2, Hy3 & +0.173 & - \\
    Cx2, Cy2  & -0.775   & - \\
    Hx4, Hx5, Hx6, Hy4, Hy5, Hy6 & +0.163 & - \\
    Cx3, Cy3  & -0.854   & - \\
    Hx7, Hx8, Hx9, Hy7, Hy8, Hy9 & +0.193 & - \\
    BrS , BrD  & -0.007 & - \\
  \hline
  \end{tabular}}
   \caption{Partial charge parameters from DFT at the b3lyp/6311-g level, and the ChelpG method \cite{breneman1990determining}, in units of $e$.}
    \end{minipage}
\end{table}

\section{ Synthesis }

\subsection{(3,3'- bis(tetradecyloxy)-[2,2'-bithiophene]-5,5'- diyl) -  \\ bis(trimethylstannane) (aT2)} 

\begin{figure}[h!]
\center
  \includegraphics[width=14cm]{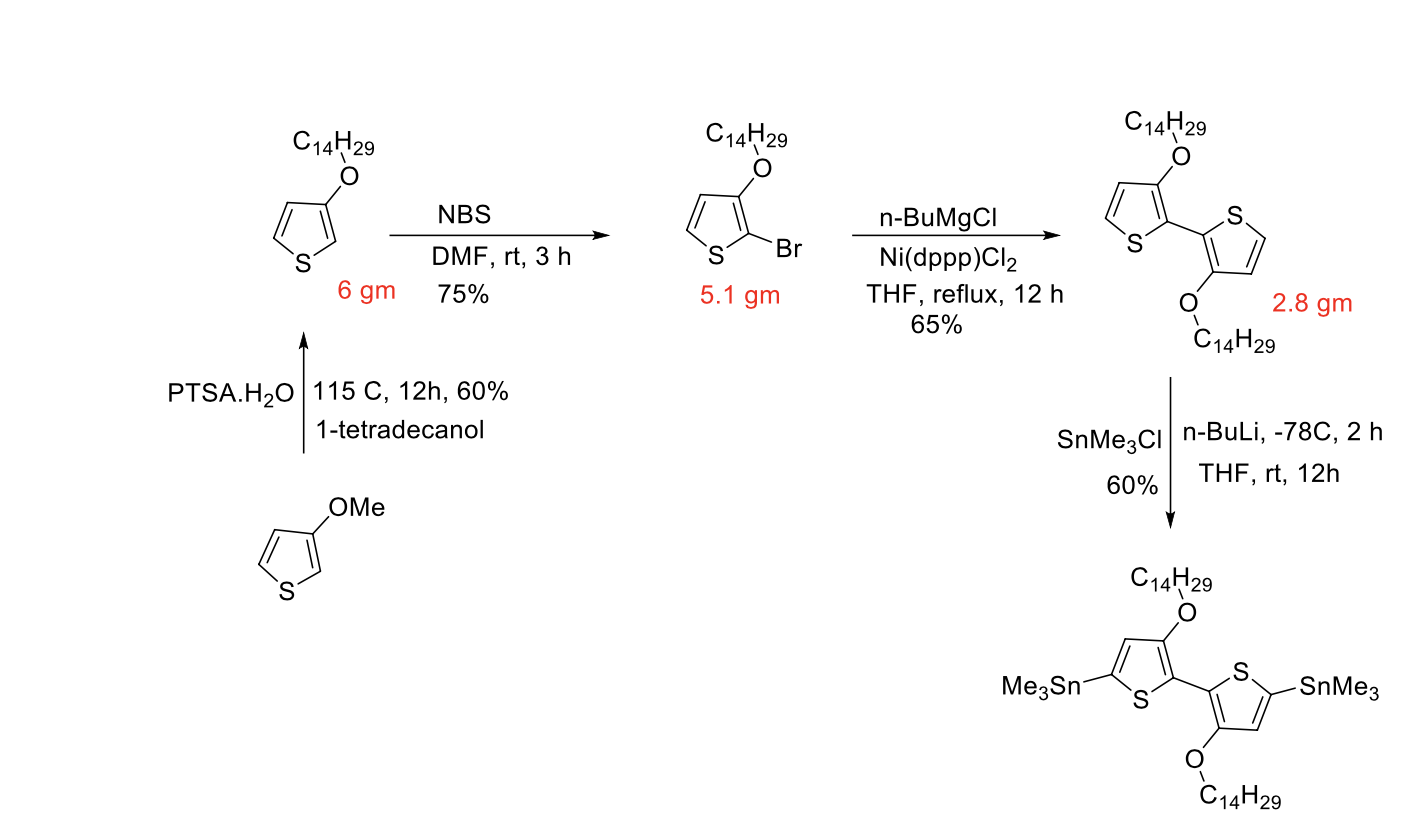}
 \caption{ Synthesis route for (3,3'-bis(tetradecyloxy)-[2,2'-bithiophene]-5,5'-diyl) bis(trimethylstannane). }
   \label{fig:Chem_Formula}
\end{figure}

A solution of 2.9 g (5.42 mmol) of 3,3'-didodecyloxy-2,2'-bithiophene \cite{guo2008conjugated} in 150mL of THF was cooled to 78° C. The resulting suspension was treated dropwise with 5.2 mL (13.0 mmol) of n-BuLi (2.5 M in hexanes) and stirred for an additional 45 minutes. The dry ice bath was then removed and the mixture was stirred at ambient temperature for 1.25 hours. The suspension was cooled to 78° C. and treated with a solution of 3.02 g (15.2 mmol) of trimethyl tin chloride in 15 mL of THF. The mixture was stirred for 90 minutes at ambient temperature. The mixture was diluted with 100 mL of ethyl acetate and washed successively with 50 mL of water and 50 mL of brine. The separated organic layer was dried over $\text{K}_2\text{CO}_3$, filtered, and concentrated in vacuo. The resulting solid was recrystallized from cold pentane to give 3.8 g (81\% yield) of 5,5'-bis(trimethylstannyl)-3,3'-bis(dodecyloxy)2,2'-bithiophene as light tan needles. m.p. 66-67$^o$ C. 1H NMR (500 MHz) $\delta$ 6.88 (s, 2H), 4.11 (t, 4H, J=6.4 Hz), 1.84 (m, 4H), 1.55- 1.30 (m, 36H), 0.88 (m, 6H), 0.36 (s, 18H).\cite{Patent1}

\subsection{poly(3,3’ditetradecoxy-bithiophene) (p(aT2))}

In a dried 5.0 mL microwave vial, 50.0 mg of 5,5'-dibromo-3,3'- bis(tetradecyloxy)-2,2'- bithiophene (66.8 $\mu$mol) and 61.3 mg of (3,3'- bis(tetradecyloxy)-[2,2'-bithiophene]-5,5'- diyl)bis(trimethylstannane) (66.8 $\mu$mol) and were dissolved in 2.5 mL of anhydrous, degassed Chlorobenzene. Pd2(dba)3 (1.2 mg, 1.36$\mu$mol) and P(o-tol)3 (1.56 mg, 5.34 $\mu$mol) were added, and the vial was sealed and heated to 100 °C for 16 h. After the polymerization has finished, the end-capping procedure was carried out. Then, the reaction mixture was cooled to room temperature and precipitated in hexane/methanol. A blue solid was formed which was filtered into a glass fibre-thimble and Soxhlet extraction was carried out with methanol, acetone, THF and chloroform. The polymer dissolved in hot chloroform. Finally, the polymer was dissolved in a minimum amount of chloroform and precipitated in methanol. The collect solid was filtered and dried under high vacuum. A blue solid w/as obtained with a yield of 66 \% (28 mg, 47.4 $\mu$mol).

GPC (DCB, 50°C) Mn = 13.8 kDa, Mw = 17.2 kDa.

\section{Methods}

\subsection{GIWAXS }

2D-Grazing Incidence Wide Angle X-ray Scattering (GIWAXS) X-ray scattering was conducted at the Stanford Synchrotron Radiation Lightsource (SSRL) at  beamline 11-3 ( with a MAR225 image-plate CCD with $73.242 \mu m \cdot 73.242 \mu m$ pixel area). Incident photon energy was 12.732 keV (0.973 ~\si{\angstrom} ) with $\sim 10^9$ photon flux.  He (g) environments for all measurements were used to minimize air scatter and beam damage to the sample. The 2D grazing-incidence sample-detector distance was 311.74 mm calibrated with a polycrystalline lanthanide hexaboride ($\text{LaB}_6$) standard at a 3.0$^o$ angle with respect to the critical angle of the calibrant. For grazing-incidence geometries, the incidence angle was set below the critical angle, to a grazing incidence of 0.1$^o$ which is above the critical angle of the underlying native oxide substrate. The data was processed using a combination of both  Nika  2D  data  reduction \cite{ilavsky2012nika}  and  homebuilt  WxDiff  Software.  The  terms  $q_{xy}$  and  $q_{z}$  denote  the component of scattering vector in-plane and out-of-plane with the substrate, respectively. Data from 2D grazing-incidence measurements were corrected for the geometric distortion introduced by a flat, plate detector and processed for subsequent analysis with the software WxDiff and GIWAXS Tool for Igor \cite{oosterhout2017mixing}. All films were prepared  on native oxide silicon substrates and measured as prepared \textit{vida supra}. 

\subsection{Molecular Dynamics} \label{section:MD}

Simulations were performed using GROMACS 2018.2 \cite{pronk2013gromacs,van2005gromacs}. Electrostatics and Van-der-Waals forces are computed using the scalable particle mesh Ewald summation \cite{hess2008gromacs}.  Hydrogen bonds are constrained using a LINCS algorithm, allowing for 2 fs time steps. 

Initial configurations were prepared using GROMACS 2018.2 and Packmol \cite{martinez2009packmol}.  Trajectories were analysed using PyMOL \cite{PyMOL}.

 Initially energy minimisation was performed on all simulations using the {\it steep} integration method, and a time step of 1 fs, until the change in energy was less than numerical accuracy.   After energy minimisation, candidate crystals were simulated in NVT at 300K for 500,000 steps at 2 fs per step. Temperature coupling is achieved through the velocity-rescale algorithm \cite{bussi2007canonical}.  Following this, a 50 ns NPT production run is performed at 300K and 1 bar using Berendsen temperature and pressure coupling \cite{berendsen1995gromacs}.  Inter molecular forces are calculated with Particle-Mesh-Ewald summations, with a 1~nm cutoff distance. 

\newpage
\subsubsection{Crystal Structure Determination and Force Field Validation}

\begin{figure}[h!]
\center
  \includegraphics[width=14cm]{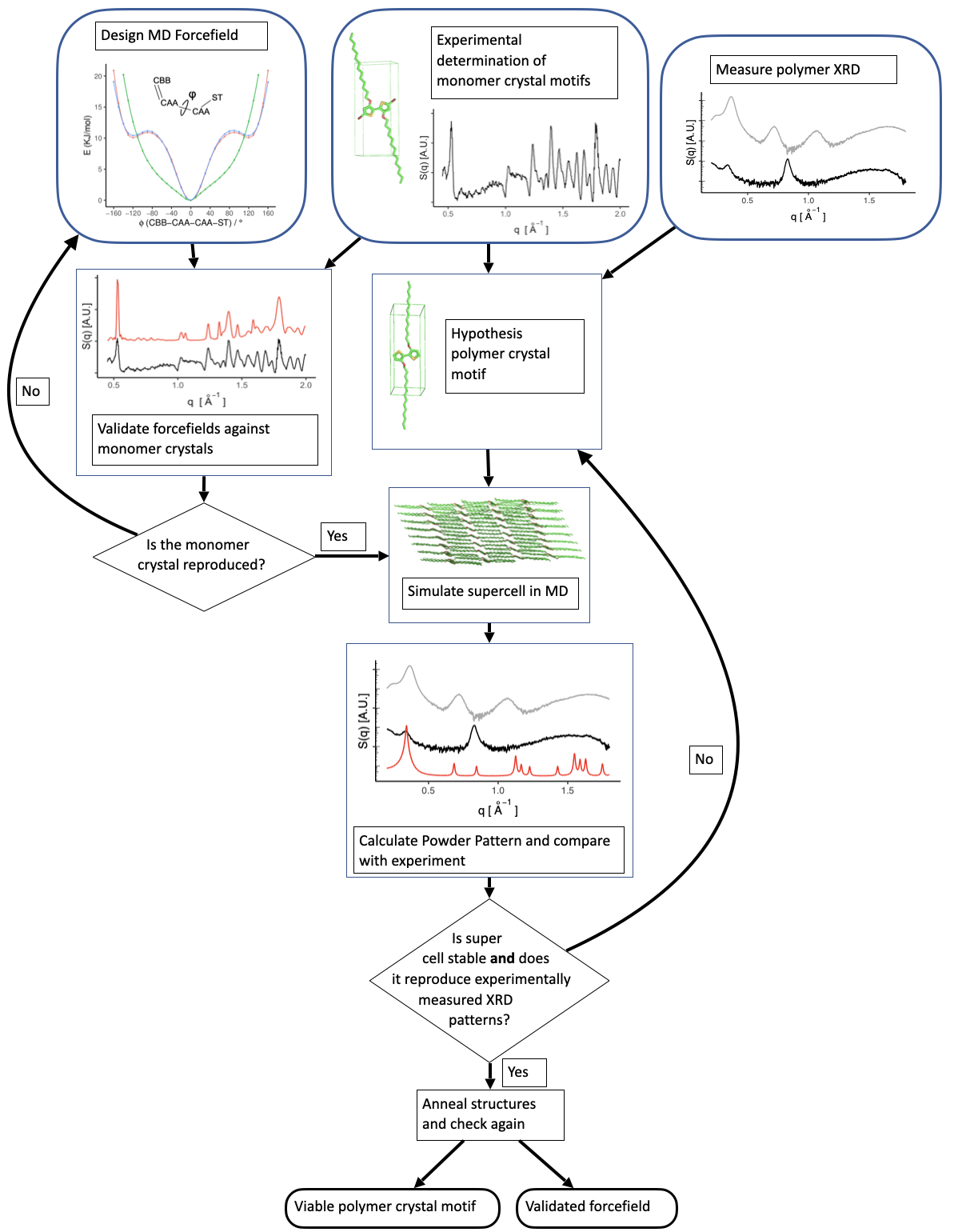}
 \caption{ Illustration of the process followed to determine the polymer unit cells and validate the MD force field.}
   \label{fig:Chem_Formula}
\end{figure}

\subsubsection{Simulated XRD Patterns }

X-ray diffraction patterns are calculated from MD in two different ways; using the Debye Equation and the Debyer package \cite{Debyer}, or using Cromer's Method \cite{hess2008gromacs}.  In both cases the q-resolution depends in the size of the simulation box, due to inverse Fourier transforms being used in calculation.  Therefore, before diffraction patterns are calculated, super cells are enlarged to contain around 800 polymers and run for 100~ps to remove the periodicity in the enlarged super cell.  Monomer crystal patterns are calculated using Debyer, p(aT2) patterns are calculated using Debyer, and p(gT2) pattern is calculated using Cromer's Method.  Scaling factors on $q$ are used where necessary to ensure proper crystal densities. For gT2 patterns, $q$ is scaled up by 2.7\%. For p(aT2), $q$ is scaled down by 14\%.

\subsubsection{Free Energy of Solvation}

Free energy of solvation for alkylated and glycolated oligomers are calculated using Bennetts Acceptance Ratio (BAR) \cite{bennett1976efficient} and 50 equally spaced $\lambda$ points. 

\subsubsection{Crystallites in Water}

After a crystallite is simulated in the solid state for 50~ns it is placed in a box with between 140,000 and 220,000 water molecules. Boxes are large enough to avoid self-interactions even if the crystallite rotates within the box. Boxes are again equilibrated according to Section \ref{section:MD} and simulated for a further 50~ns in NPT as a production run. 

\subsubsection{Wet Bulk Crystals}

Wet bulk crystals are simulated by taking dry crystal structures and inserting 20 water molecules per polymer inside the crystal. Systems are then equilibrated according to Section \ref{section:MD}. They are further simulated in NPT for 50ns.  

\subsubsection{Metadynamics}

Metadynamics simulations were performed on a trimer of p(gT2) in a $6 \text{nm}^3$ box of water containing one Na$^{+}$ ion and one Cl$^{-}$ counterion. The box was equilibrated using the same parameters and methods outlined in Section \ref{section:MD}.   

Following equilibration the free energy surfaces (FES) were determined using well-tempered metadynamics simulations \cite{Laio12562, PhysRevLett.100.020603} using Plumed 2.4.1 \cite{Bonomi2009}. Two collective variables (CVs) were used to drive the simulation, $R_1$ and $R_2$, biasing the Na$^{+}$ interaction with side chains on the top of the oligomer and bottom of the oligomer respectively. $R_1$ and $R_2$ are defined as the minimum distance between the Na$^{+}$ and the glycol oxygen centres of mass (see Figure \ref{fig:CVs}).

Hills were deposited along these two CVs every 200 steps using an initial hill height of 3~kJ/mol and a bias factor of 10. The gaussian sigma width for each CV was set to 0.05~nm. The metadynamics simulation was run for a total of $4 \mu s$ allowing for a thorough exploration of the CV space.  

In order to allow this simulation to converge in accessible time-scales the CV space was limited using two external 
potentials as upper walls ($V_{1}$ and $V_{2}$), limiting the values of each CV. These are given by
$$V_{1} = 150(x_{1} - 1.6)^2$$ 
$$V_{2} = 150(x_{2} - 3.42)^2$$ 
where 
$$x_{1} = R_2 - (0.467 \cdot R_1)$$
and 
$$x_{2} = (2.14 \cdot R_1) - R_2.$$
The potential $V_{1}$ was active on the system when $x_1 > 1.6$ and $V_{2}$ was active when $x_2 > 3.42.$ Preliminary  metadynamics simulations (not shown here) exploring the CV between 0 and 3.5 in each CV indicated that the potentials above do not occlude any interactions between the binding sites and the Na$^{+}$ ion. 

To make use of parallelisation in our calculation, an initial simulation was run for $1 \mu s$ to generate 200 initial configurations, equally spaced by $5 ns$. Each configuration was used as a starting point for a separate walker, which was then run for $20 ns$ each sharing the same free energy surface, leading to an effective speed of $2 \mu s$ per day calculation speed \cite{raiteri2006efficient}. 

Convergence of the FES was assured through three measures. The chemistry, and the way the CVs are defined, should result in a symmetric FES along $R_2=R_1$. Symmetry of the FES is assessed by observation during the simulation.  Secondly, the reduction of large hills being deposited, indicating that the free energy surface has mostly filled up.  Thirdly, the binding energies are monitored during the simulation as shown in Figure \ref{fig:binding_t_dep}. When the binding energies stop changing with time or begin to oscillate around a value, the simulation has converged. 

Minimum energy transition pathways are calculated using a nudged elastic band method \cite{henkelman2000climbing, henkelman2000improved}.  

\begin{figure}[h!]
\center
  \includegraphics[width=8cm]{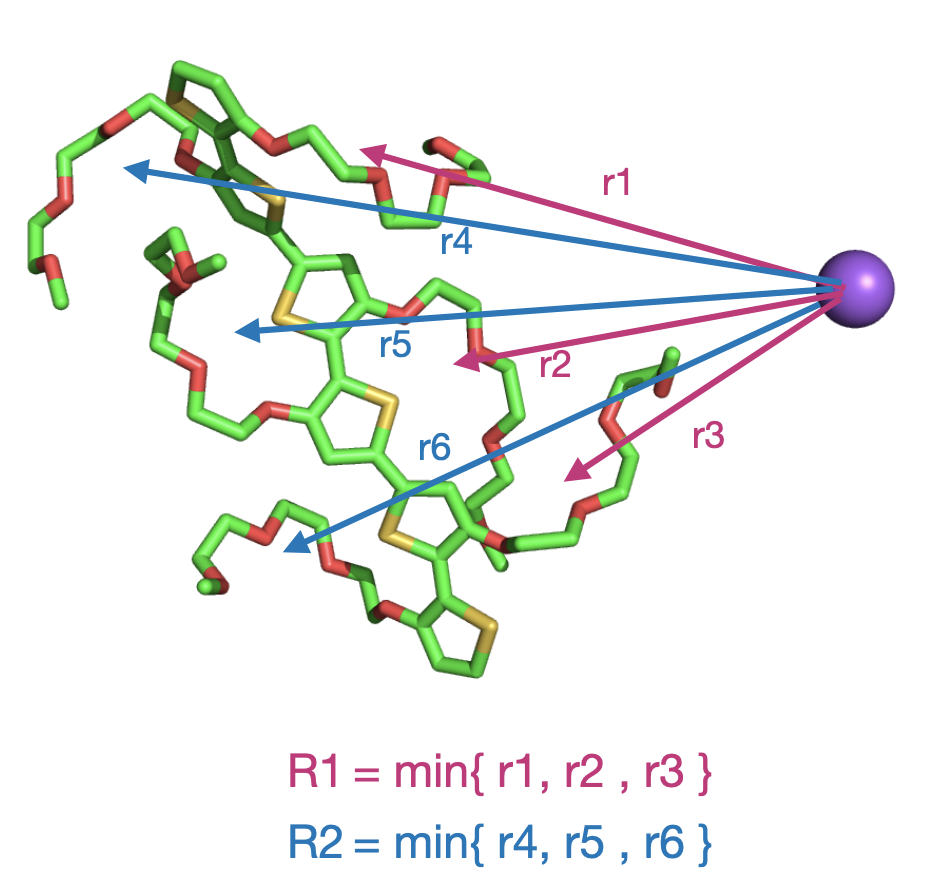}
 \caption{ Illustration of how the CVs are defined in the metadynamics simulation }
   \label{fig:CVs}
\end{figure}

\clearpage
\section{Force Field Validation}

\begin{figure}[h!]
\center
  \includegraphics[width=11cm]{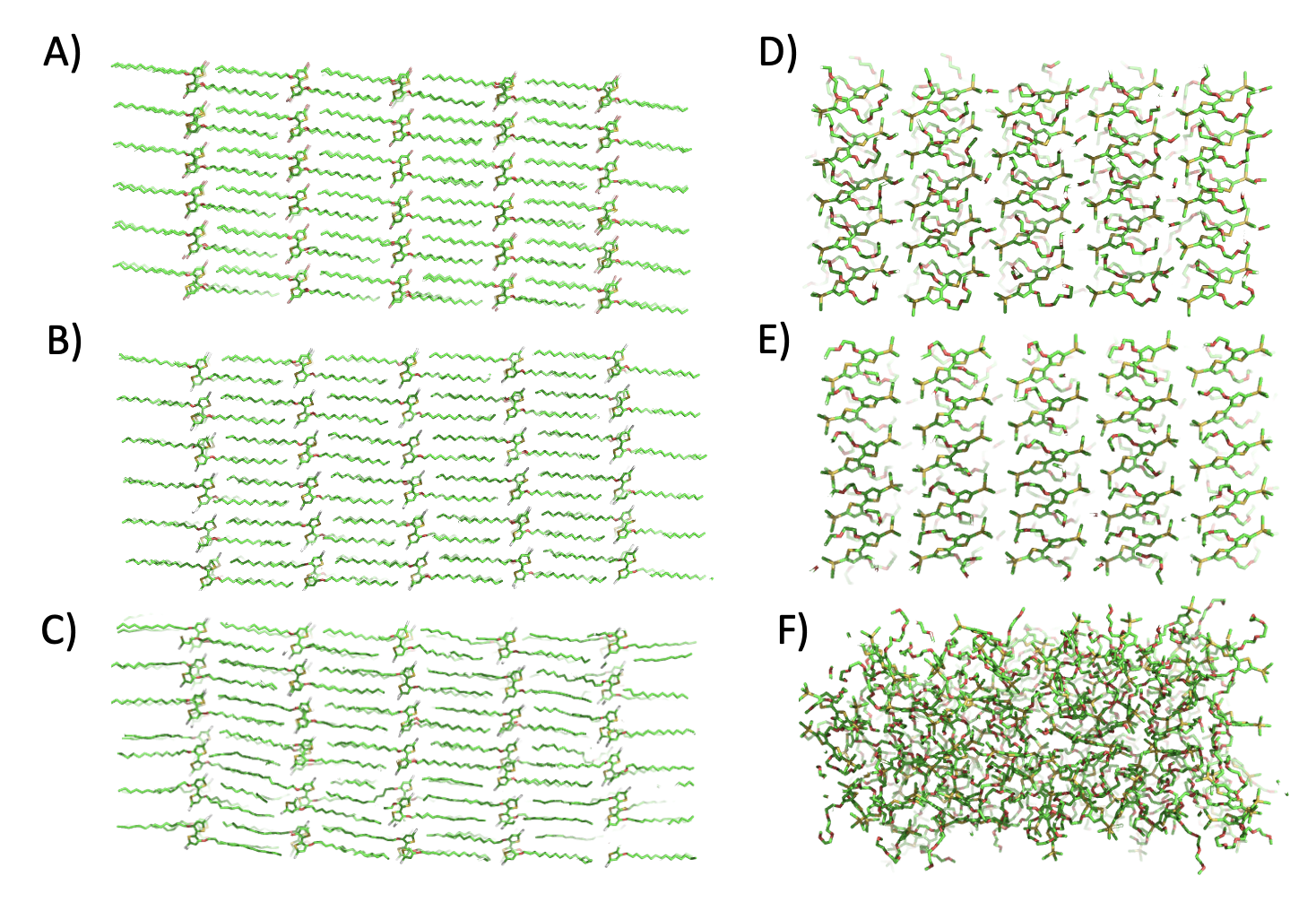}
 \caption{ Snapshots of aT2 (left panel) and gT2 (right panel) crystal after annealing at 300K (top), 350K (middle) and 400K (bottom) for 1ns followed by cooling at a rate of 10K per ns back to room temperature.     }
   \label{fig:binding_t_dep}
\end{figure}

\begin{figure}[h!]
\center
  \includegraphics[width=11cm]{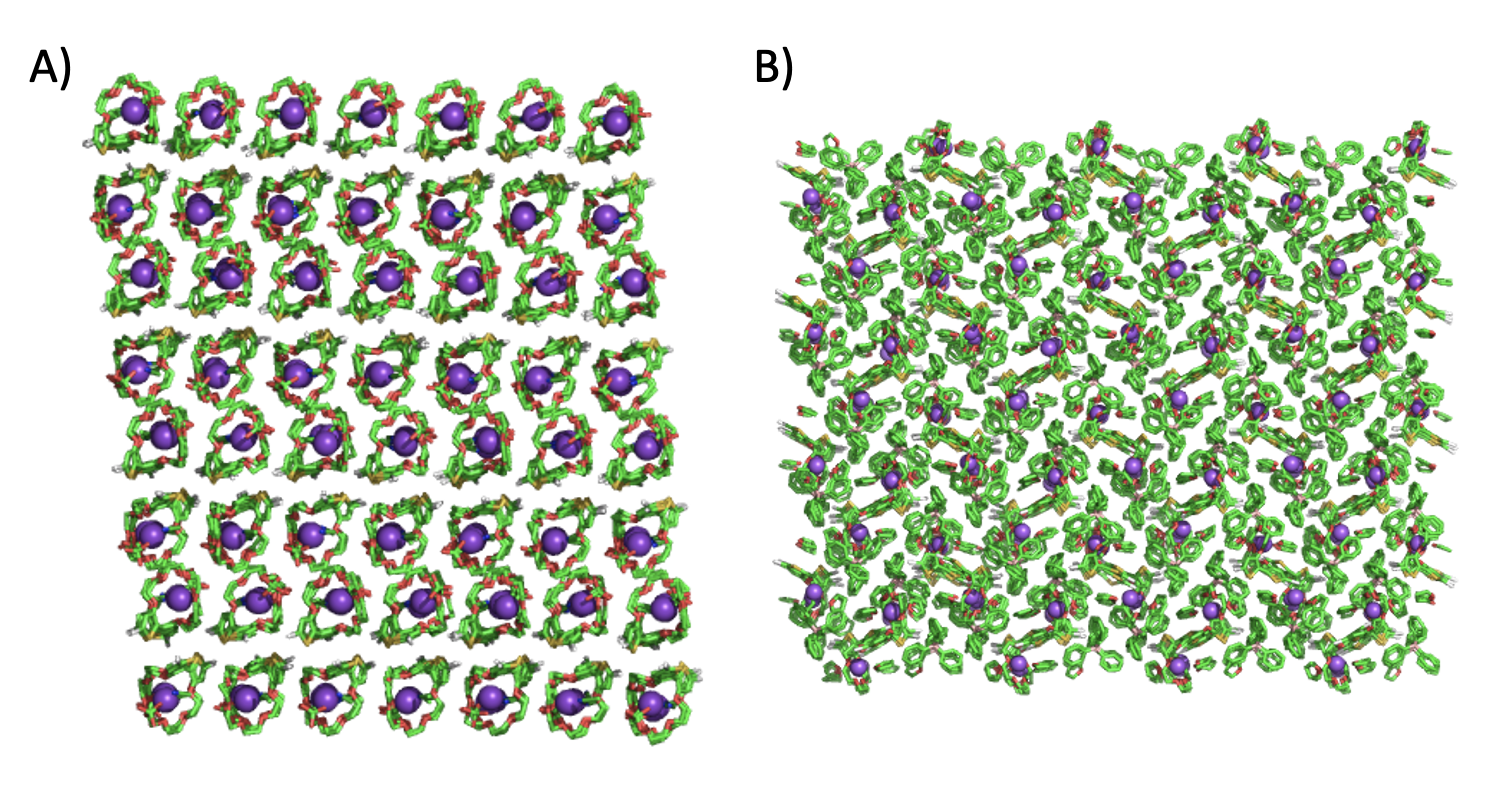}
 \caption{ Snapshots of a) (17‐Crown‐5)T2:NaBPh4 and b) (20‐Crown‐6)T2:KClO4 oligomer-ion co-crystals presented by A. Giovannitti \cite{Giovannitti2016b} used for validation of charge interactions. Snapshots are after 50ns of simulation in MD at room temperature.  }
   \label{fig:binding_t_dep}
\end{figure}

\clearpage
\section{Additional Results}

\subsection{Crystallography}

\subsubsection{aT2}

\begin{figure}[h!]
\center
  \includegraphics[width=11cm]{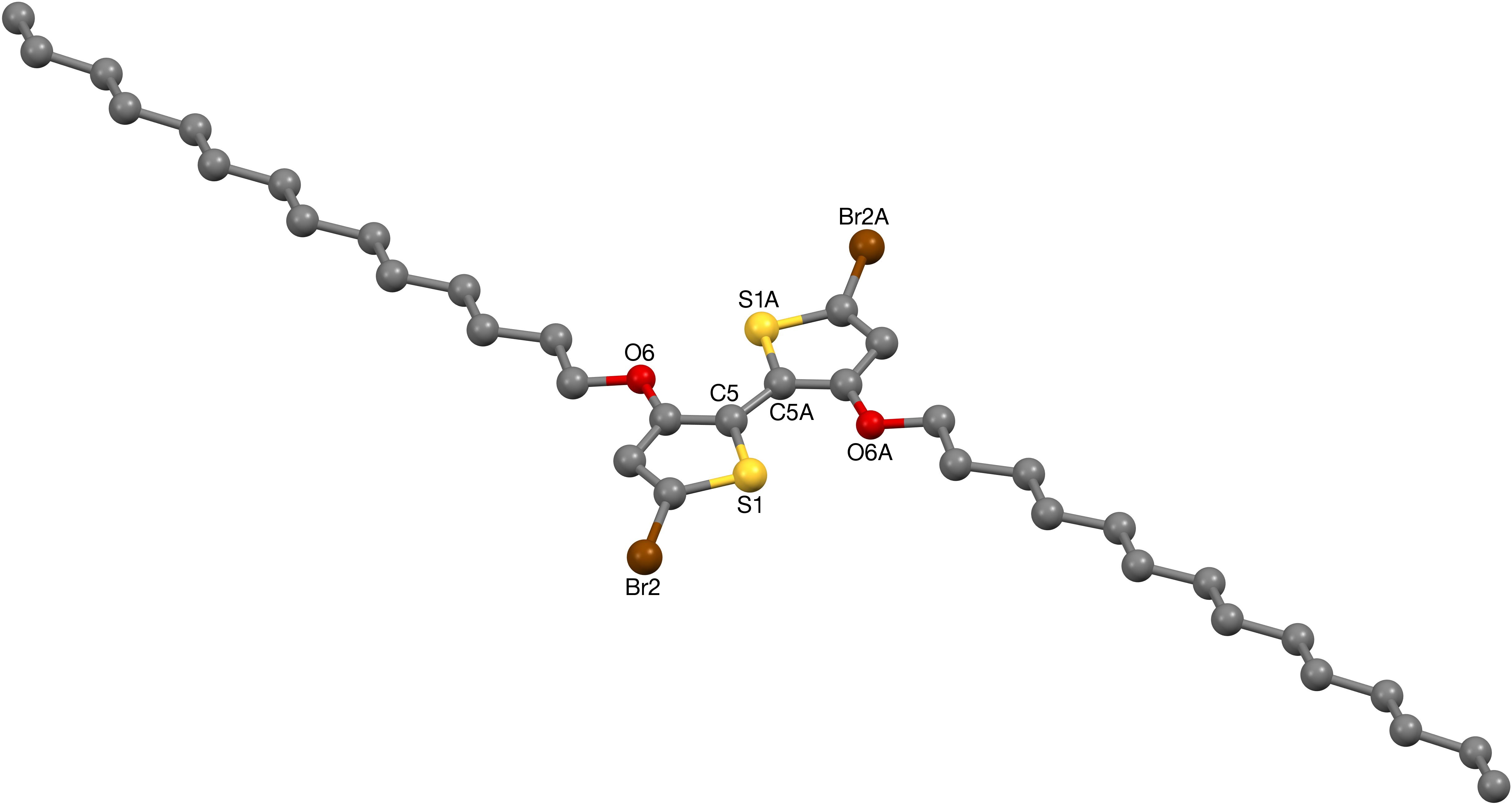}
 \caption{ The crystal structure of the Ci-symmetric molecule aT2 (50\% probability ellipsoids). }
   \label{fig:binding_t_dep}
\end{figure}

\textit{Crystal Data for aT2:} $\text{C}_{36} \text{H}_{60} \text{Br}_2 \text{O}_2 \text{S}_2$, $M = 748.78,$ triclinic, $P-1$ (no. 2), $a = 4.1652(2),$ $b = 9.2979(4),$ $c = 23.8056(13) \, \si{\angstrom},$ $\alpha = 91.863(4),$ $\beta = 91.478(4),$ $\gamma = 96.453(4)^o,$ $V = 915.20(8) \, \si{\angstrom} ^3,$ $Z = 1 \, [C_i \text{ symmetry}],$ $D_c = 1.359  \text{ g cm}^{-3},$ $\mu(\text{Cu-K}\alpha) = 4.089 \text{ mm}^{-1},$ $T = 173 \, K,$ colourless thin platy needles,  Agilent Xcalibur PX Ultra A diffractometer, 3490 independent measured reflections ($R_\text{int}$ = 0.0309), $F^2$ refinement \cite{bruker1998saint, sheldrick2008short}, $R_1(\text{obs}) = 0.0433$, $wR_2(\text{all}) = 0.1141$, 3056 independent observed absorption-corrected reflections [$|F_0| > 4\sigma (|F_0|)$, completeness to $\theta_\text{full}(67.7^o) = 98.0\%$], 191 parameters. CCDC 2116774.

The structure of aT2 sits across a centre of symmetry at the middle of the C5–C5A bond.

\clearpage
\subsubsection{gT2}

\begin{figure}[h!]
\center
  \includegraphics[width=8cm]{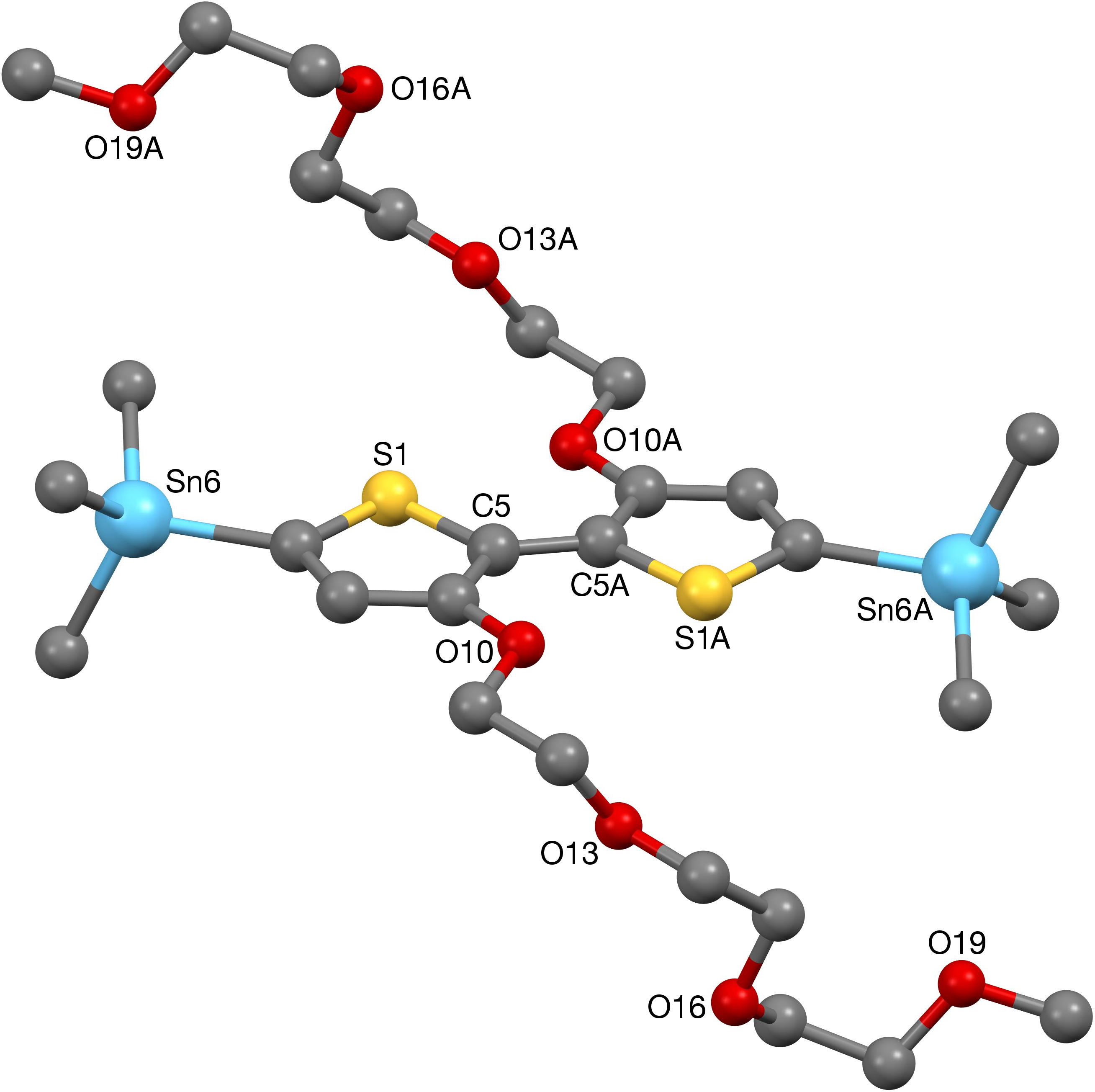}
 \caption{ The crystal structure of the Ci-symmetric molecule gT2 (50\% probability ellipsoids). }
   \label{fig:binding_t_dep}
\end{figure}

\textit{Crystal Data for gT2:} $\text{C}_{28} \text{H}_{50}  \text{O}_8 \text{S}_2 \text{Sn}_2,$ $M = 816.18,$ monoclinic, $P2_1/n$ (no. 14), $a = 7.00588(12),$ $b = 14.0253(3),$ $c = 18.0156(3) \, \si{\angstrom},$ $\beta = 97.7113(16)^o,$ $V = 1754.20(6) \, \si{\angstrom} ^3,$ $Z = 2 \, [C_i  \text{ symmetry}],$ $D_c = 1.545 \text{ g cm}^{-3},$ $\mu(\text{Cu-K}\alpha) = 12.783 \text{ mm}^{-1},$ $T = 173 \, K,$ colourless blocky needles,  Agilent Xcalibur PX Ultra A diffractometer; 3378 independent measured reflections ($R_\text{int} = 0.0336$), $F^2$ refinement \cite{bruker1998saint, sheldrick2008short}, $R_1(\text{obs}) = 0.0323,$ $wR_2(\text{all}) = 0.0858,$ 2892  independent observed absorption-corrected reflections [$|F_0| > 4\sigma (|F_0|),$ completeness to $\theta_\text{full} (67.7^o) = 98.2\%],$ 186 parameters. CCDC 2116774.

The structure of gT2 sits across a centre of symmetry at the middle of the C5–C5A bond.

\clearpage
\subsection{GIWAXS}

\begin{figure}[h!]
\center
  \includegraphics[width=11cm]{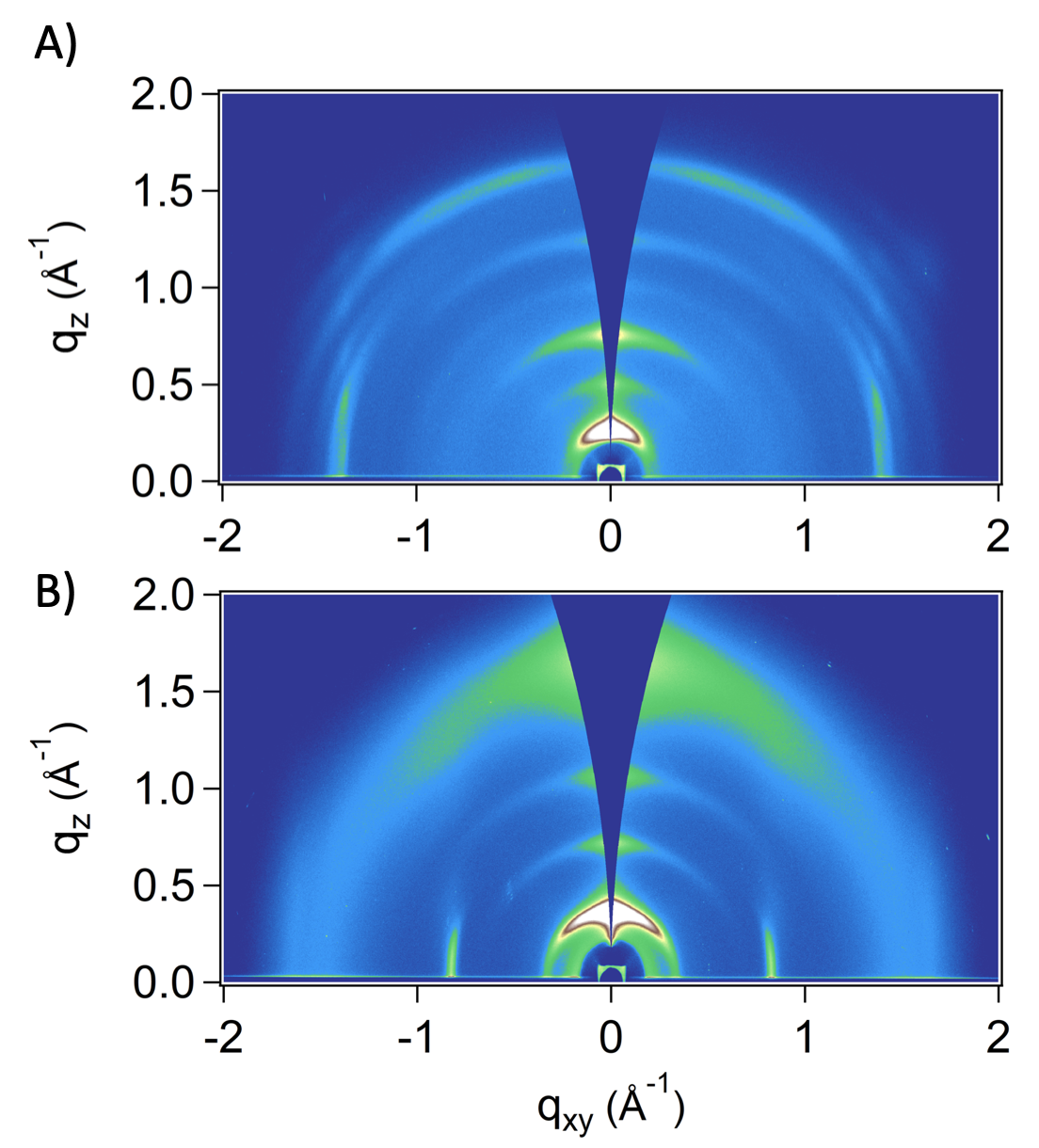}
 \caption{ Measured 2D GIWAXS pattern for A) p(aT2) and B) pg(T2).  }
   \label{fig:binding_t_dep}
\end{figure}

\begin{figure}[h!]
\center
  \includegraphics[width=9cm]{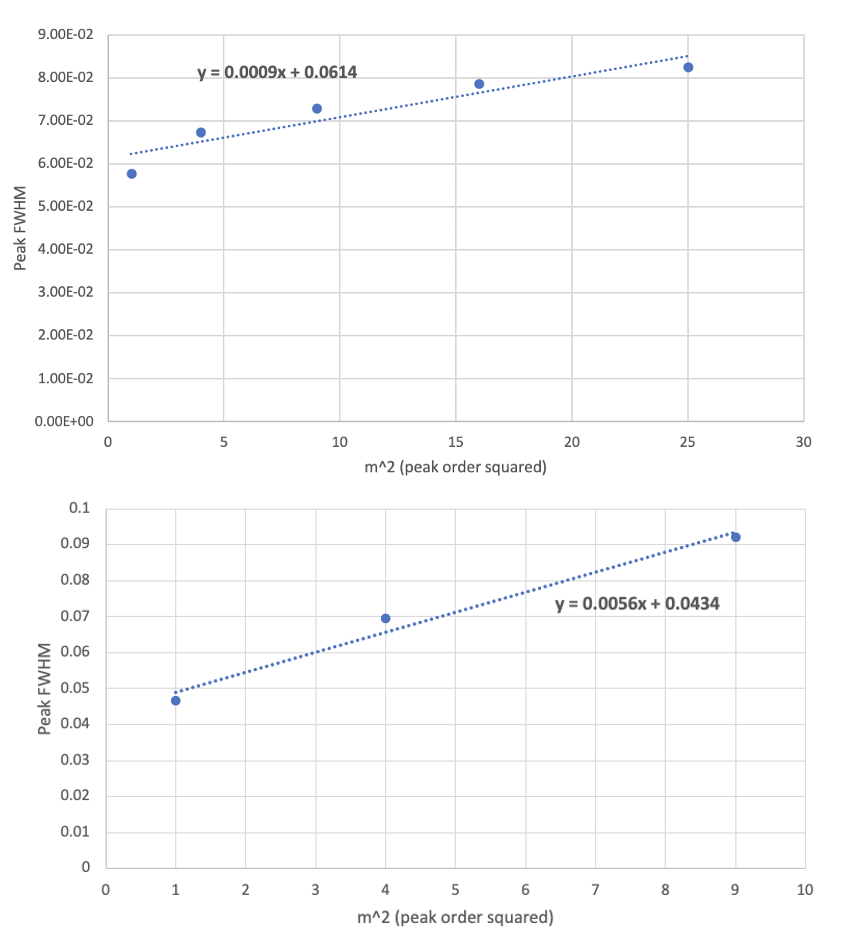}
 \caption{ Extraction of coherence lengths from fitting peak width against peak order for the lamellar peaks ((h 0 0) family).  Line fitting gives a coherence length of 10.2~nm (around 4 planes) and 14.5~nm (around 8 planes) for p(aT2) (top) and p(gT2) (bottom).   }
   \label{fig:binding_t_dep}
\end{figure}

\clearpage
\subsection{Nuclear Magnetic Resonance (NMR)}

\begin{figure}[h!]
\center
  \includegraphics[width=18cm]{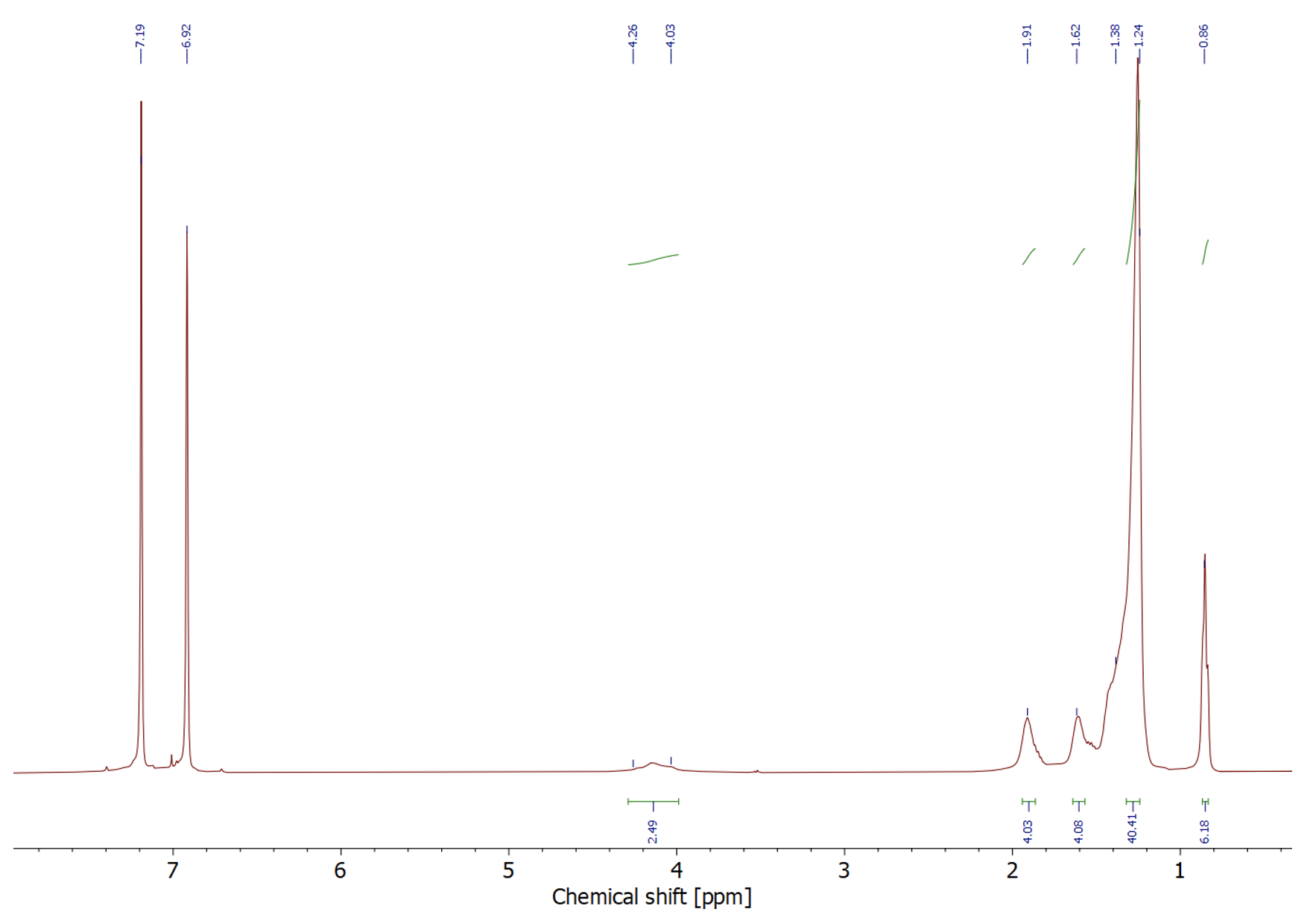}
 \caption{ $^1H$ NMR (1,4-dichlorobenzene-$d_4$) ((500 MHz, 373 K): 4.26 - 4.03 (m, 4 H, $\text{OCH}_2$), 1.96 - 1.82 m (m, 4H), 1.66 - 1.52 (m, 4H), 1.42 - 1.20 (m, 40H signal overlapping with $\text{H}_2\text{O}$ peak). The protons from the dialkoxybithiophene unit are overlapping with the solvent peaks of 1,4-dichlorobenzene-$\text{d}_4$) and cannot be resolved.  }
   \label{fig:binding_t_dep}
\end{figure}

\subsection{Photoelectric Spectroscopy in Air (PESA)}

PESA measurements indicate p(aT2) has an ionisation potential of 4.546 eV.

\subsection{Crystal Structure Factors }

Theoretical structure factors for perfect crystals are obtained from Mercury \cite{macrae2020mercury}.  

\begin{figure}[h!]
\center
  \includegraphics[width=12cm]{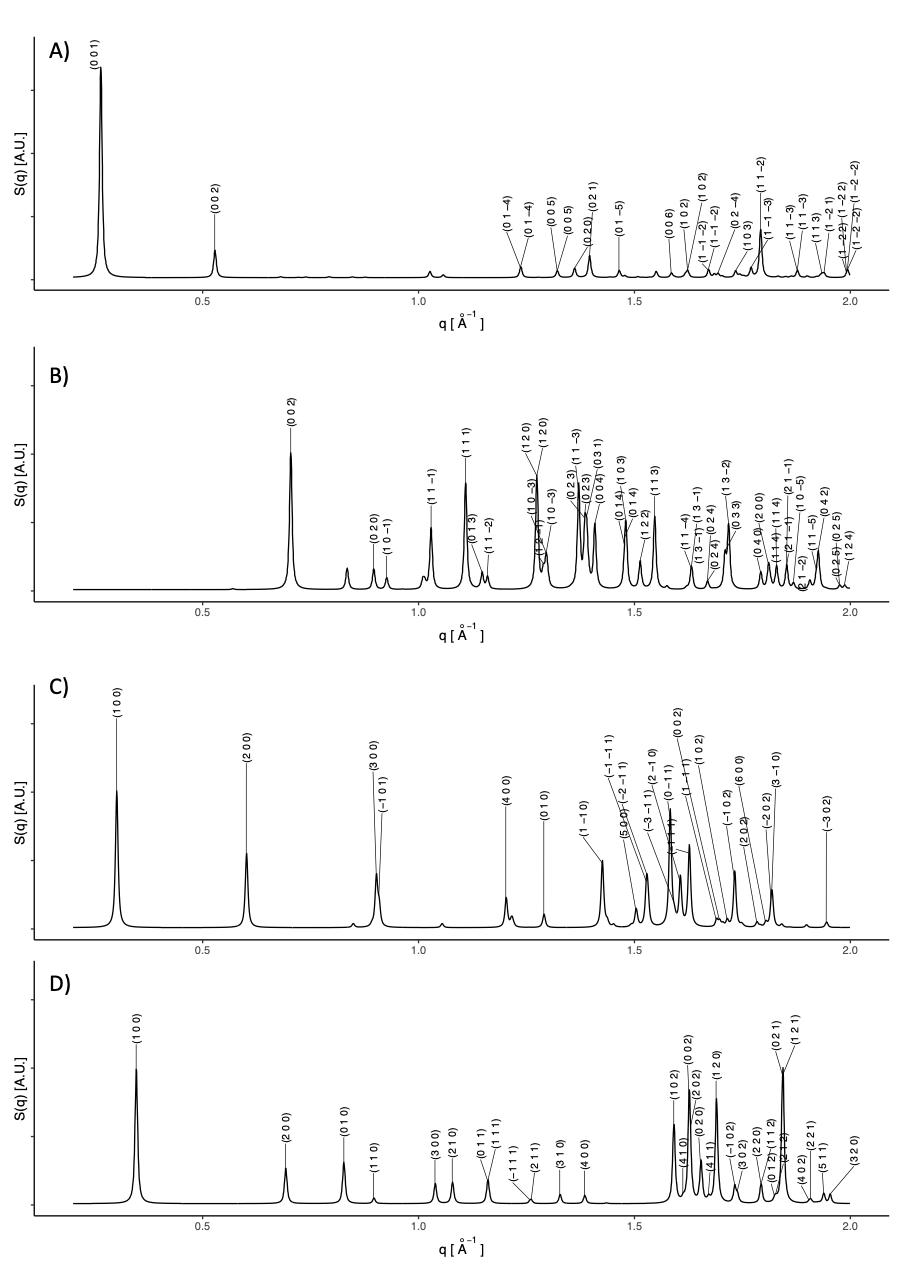}
 \caption{ Structure factors for a) aT2, b) gT2, c) paT2 and d) pgT2 crystals. }
   \label{fig:binding_t_dep}
\end{figure}

\begin{figure}[h!]
\center
  \includegraphics[width=14cm]{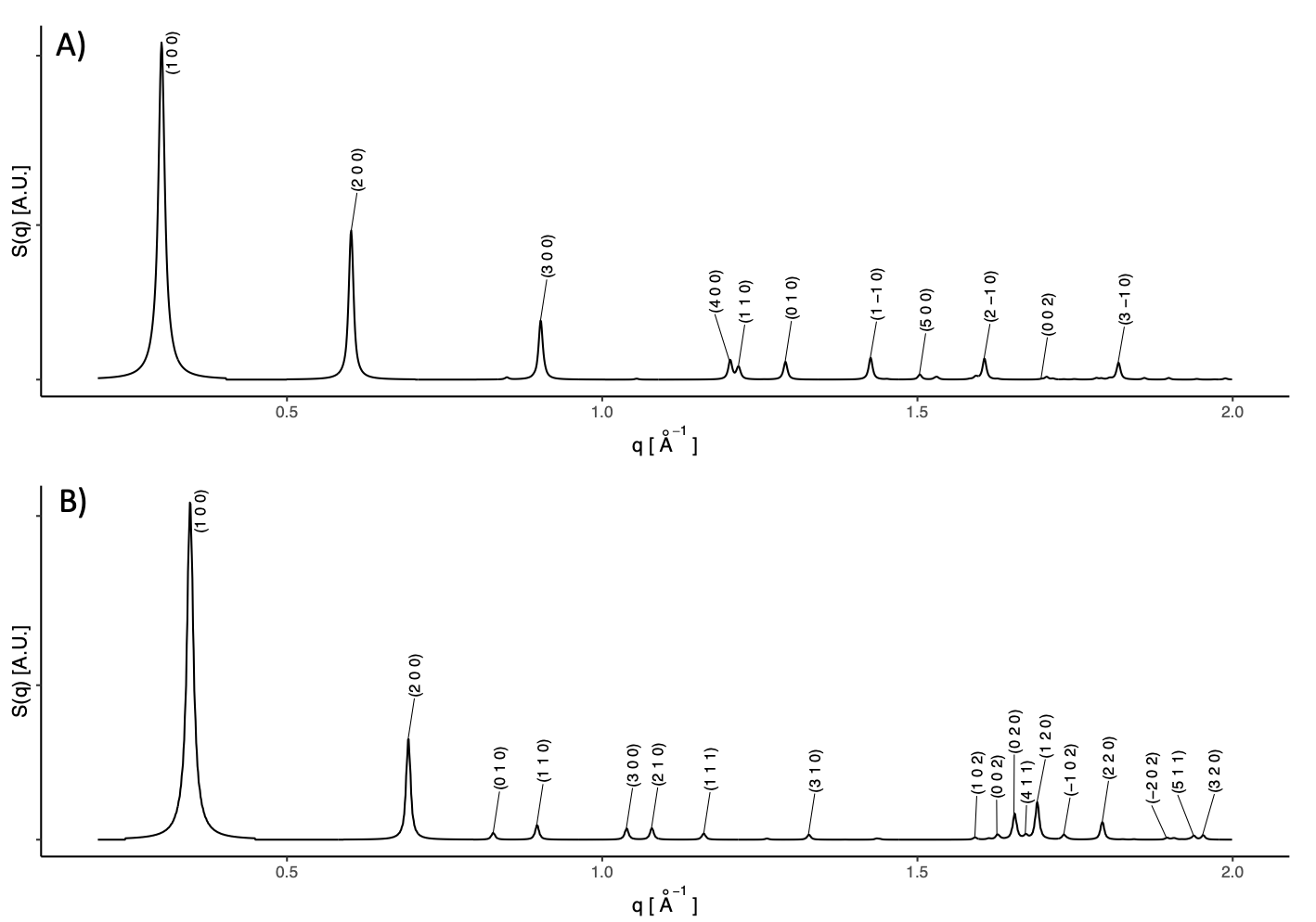}
 \caption{ Structure factors polymer backbones in a) p(aT2), b) p(gT2) crystals. }
   \label{fig:binding_t_dep}
\end{figure}

\clearpage
\subsection{ Simulated XRD Patterns }

\begin{figure}[h!]
\center
  \includegraphics[width=18cm]{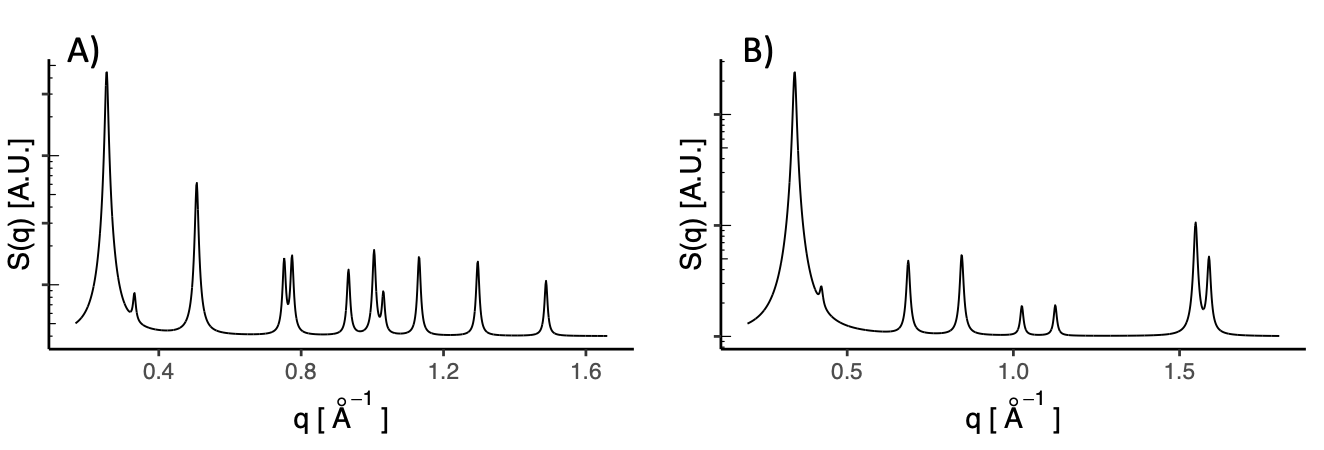}
 \caption{ Simualted XRD Patterns for a) paT2 and b) pgT2 backbones   }
   \label{fig:binding_t_dep}
\end{figure}

\clearpage
\subsection{ Crystallites in Water }

\begin{figure}[h!]
\center
  \includegraphics[width=15cm]{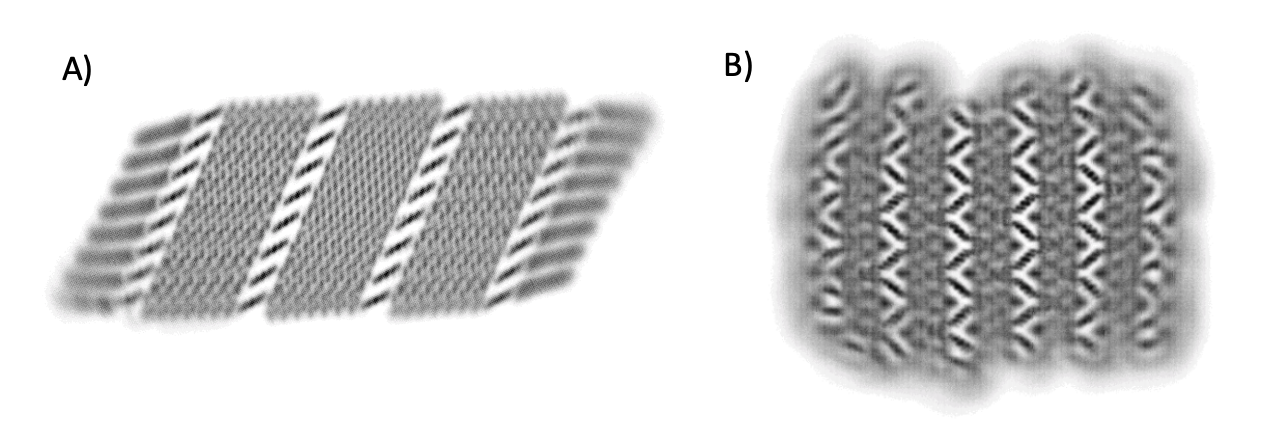}
 \caption{ Density plots of polymer atoms for the a) p(aT2) and b) p(gT2) polymers when immersed in a water bath. In both cases backbones remain ordered, however in the pgT2 crystal, the side chains have become disordered, shown by the smearing of the density from that part of the crystallite. }
   \label{fig:Chem_Formula}
\end{figure}

\begin{figure}[h!]
\center
  \includegraphics[width=15cm]{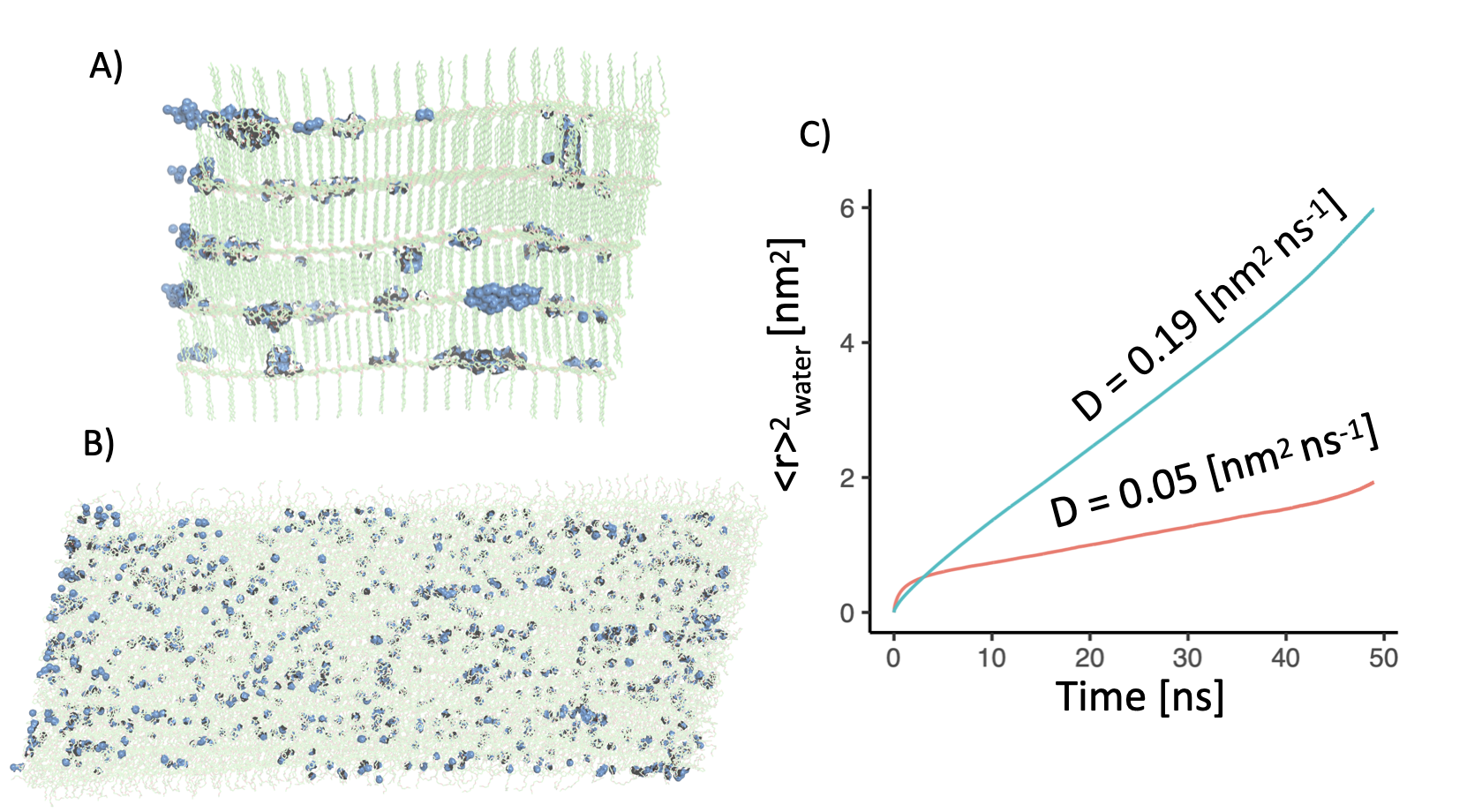}
 \caption{ Snapshots of bulk wet a) p(aT2) and b) p(gT2) crystals after 50~ns of simulation time, with c) the simulated water diffusion coefficients, D, for p(aT2) (red line) and p(gT2) (blue line). When compared to the bulk self-diffusion coefficient of water ($2.29 \, \text{nm}^2 \text{ns}^{-2}$), when in p(gT2) water diffuses an order of magnitude slower, whilst when in p(aT2), water diffuses 3 orders of magnitude slower.  }
   \label{fig:Chem_Formula}
\end{figure}

\subsection{Ion Chelation for Species with Different Side Chain Lengths}

\begin{figure}[h!]
\center
  \includegraphics[width=14cm]{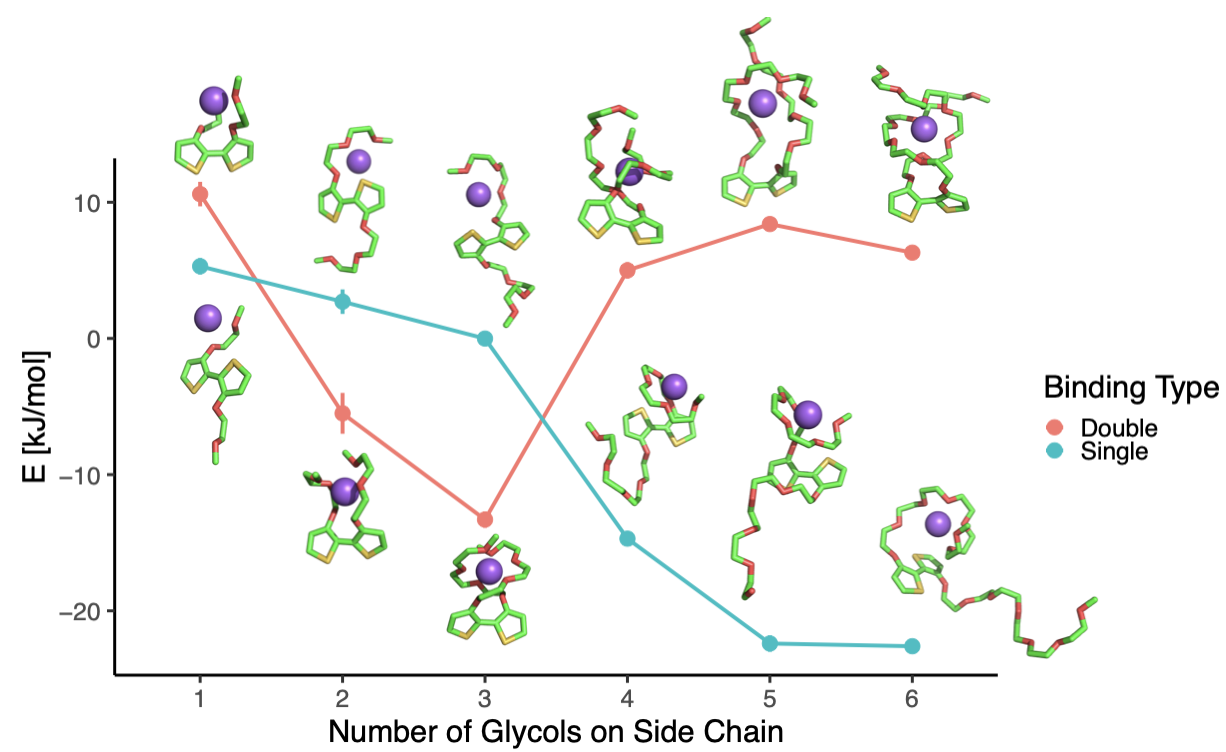}
 \caption{ Single and double bound chelation energies as a function of the length of the glycol side chain. Single monomers shown for clarity. }
   \label{fig:CVs}
\end{figure}

\begin{figure}[h!]
\center
  \includegraphics[width=10cm]{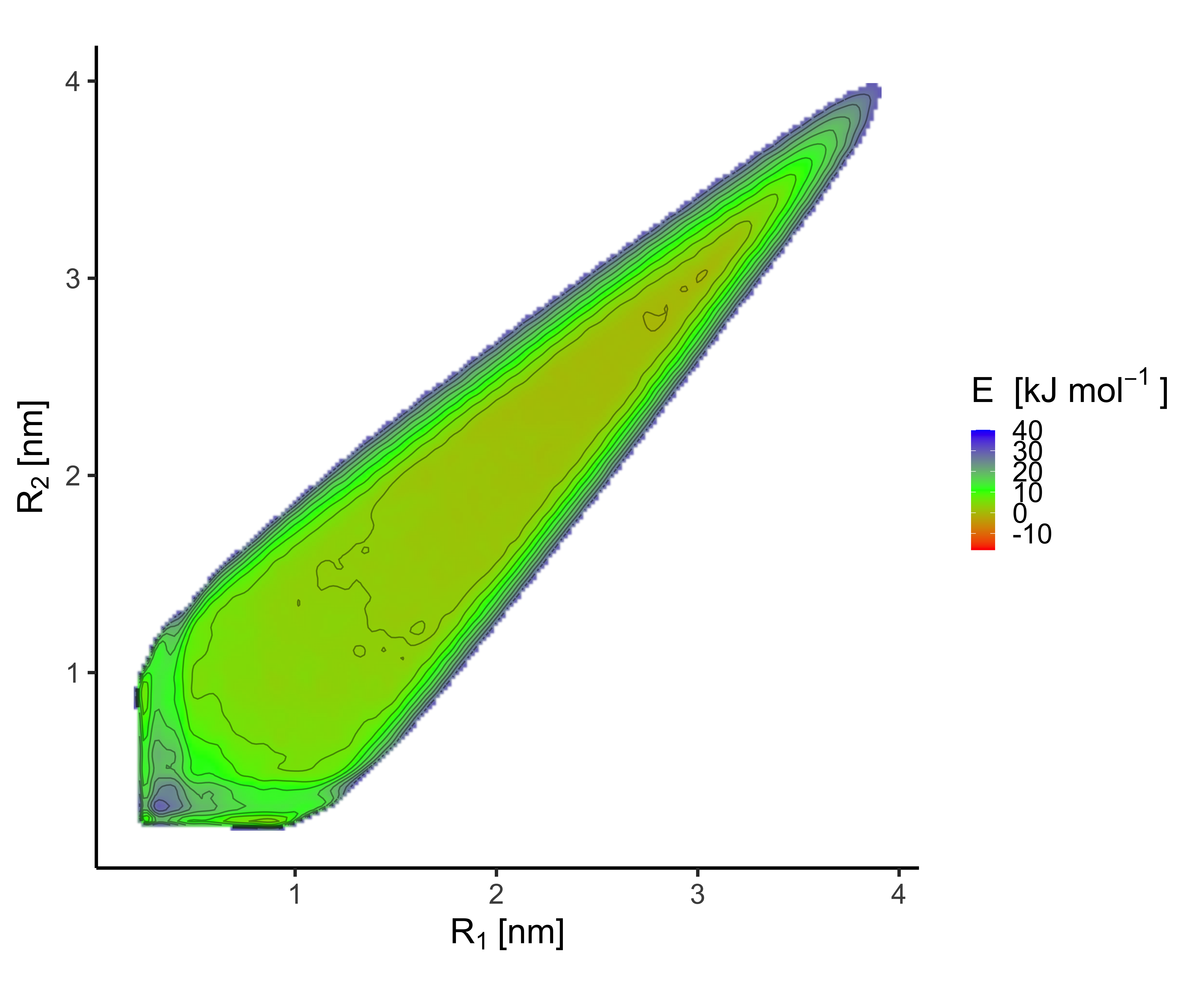}
 \caption{ Free Energy Surface for a Na$^{+}$ ion and mono-ethylene glycol side chains.   }
   \label{fig:CVs}
\end{figure}

\begin{figure}[h!]
\center
  \includegraphics[width=10cm]{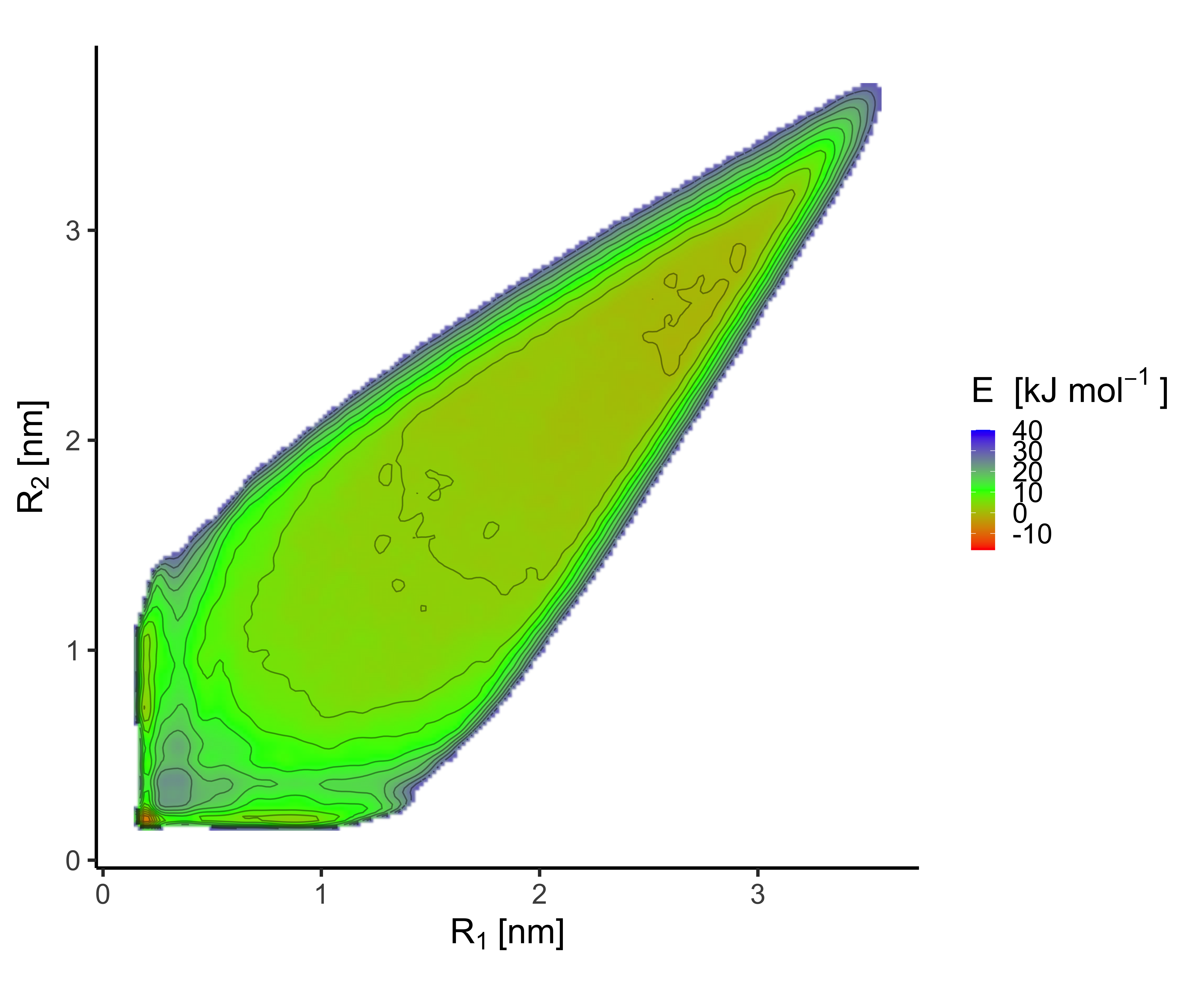}
 \caption{ Free Energy Surface for a Na$^{+}$ ion and bi-ethylene glycol side chains.   }
   \label{fig:CVs}
\end{figure}

\begin{figure}[h!]
\center
  \includegraphics[width=10cm]{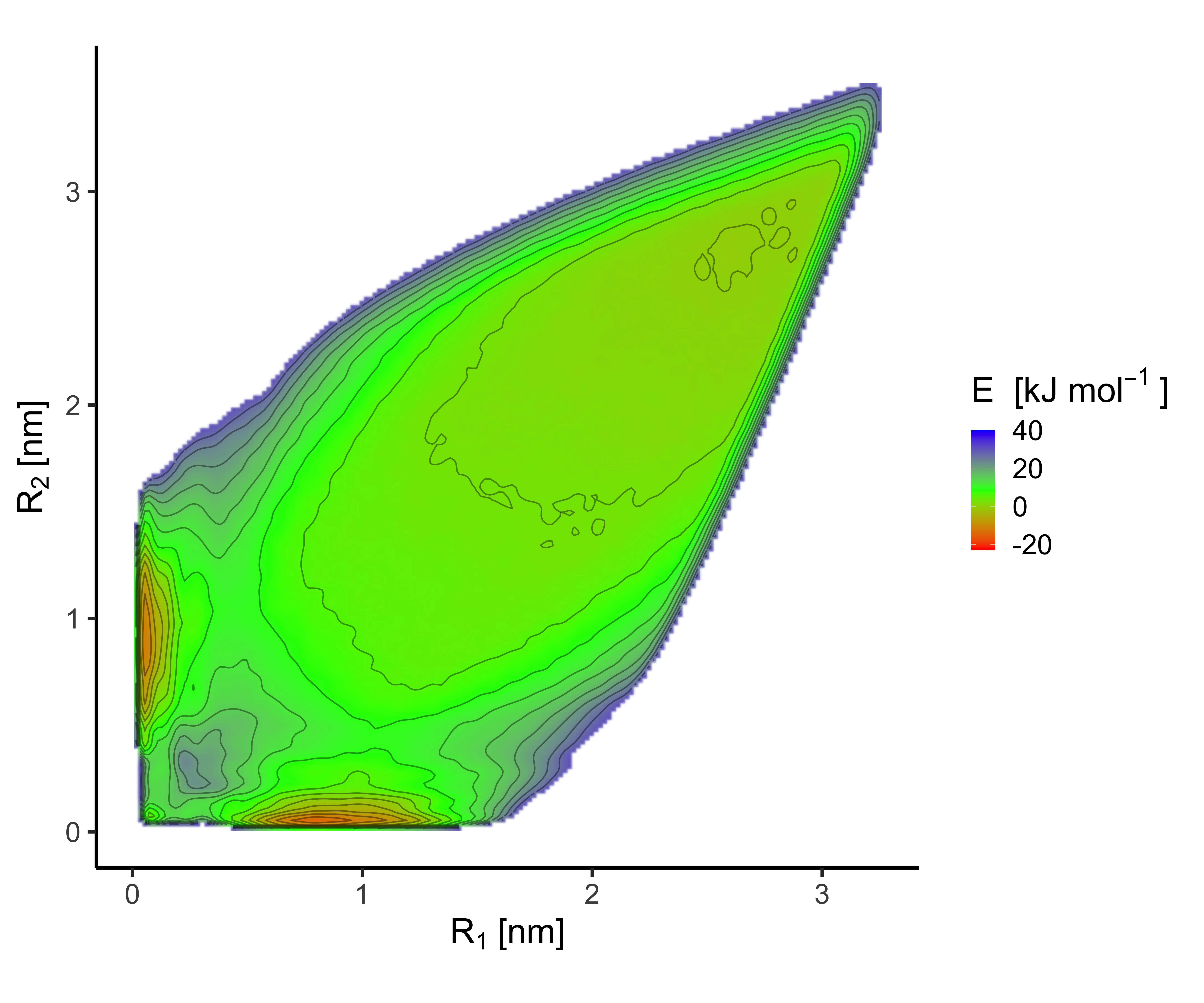}
 \caption{ Free Energy Surface for a Na$^{+}$ ion and tetra-ethylene glycol side chains.   }
   \label{fig:CVs}
\end{figure}

\begin{figure}[h!]
\center
  \includegraphics[width=10cm]{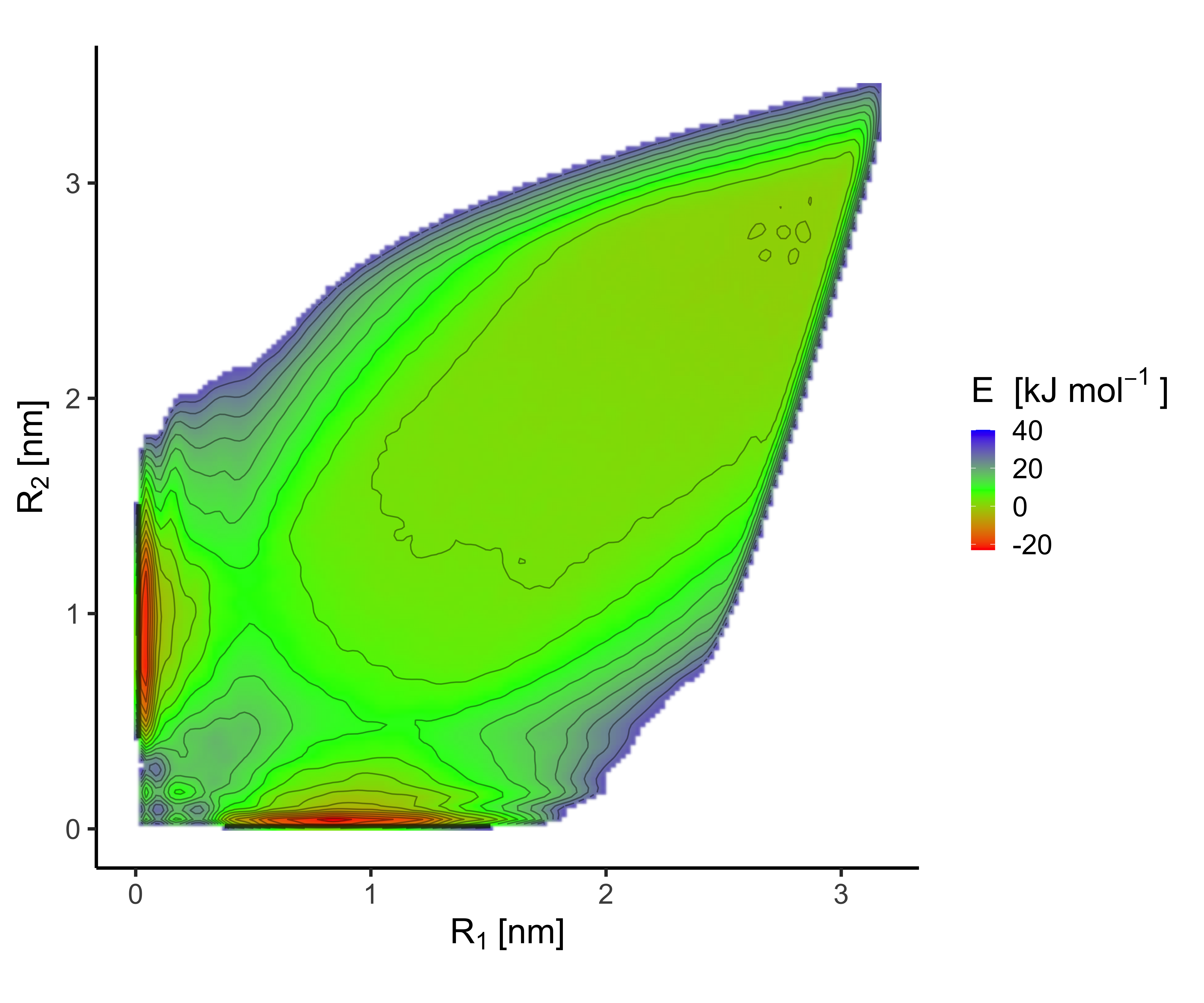}
 \caption{ Free Energy Surface for a Na$^{+}$ ion and penta-ethylene glycol side chains.   }
   \label{fig:CVs}
\end{figure}

\begin{figure}[h!]
\center
  \includegraphics[width=10cm]{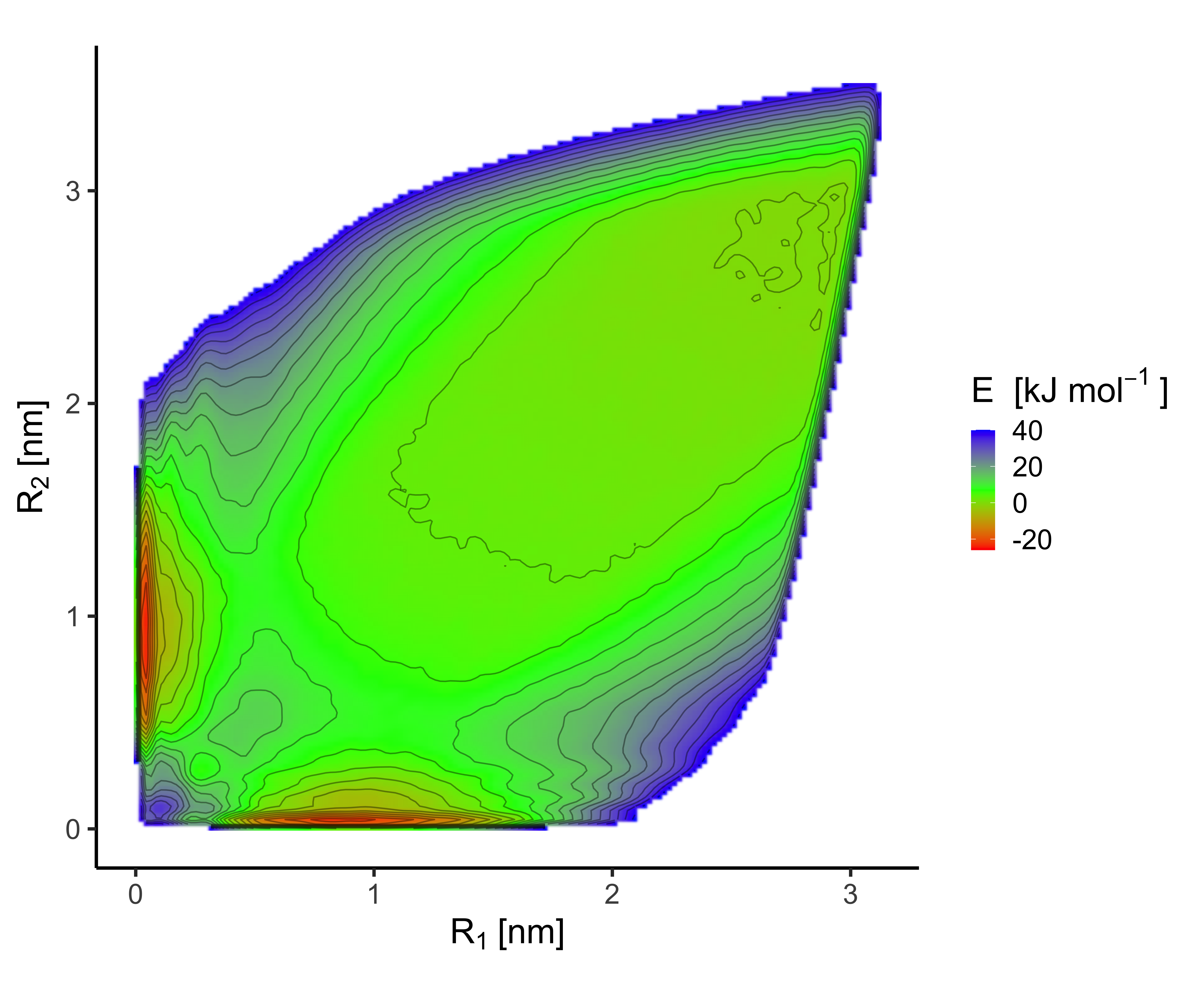}
 \caption{ Free Energy Surface for a Na$^{+}$ ion and hexa-ethylene glycol side chains.   }
   \label{fig:CVs}
\end{figure}

\begin{figure}[h!]
\center
  \includegraphics[width=10cm]{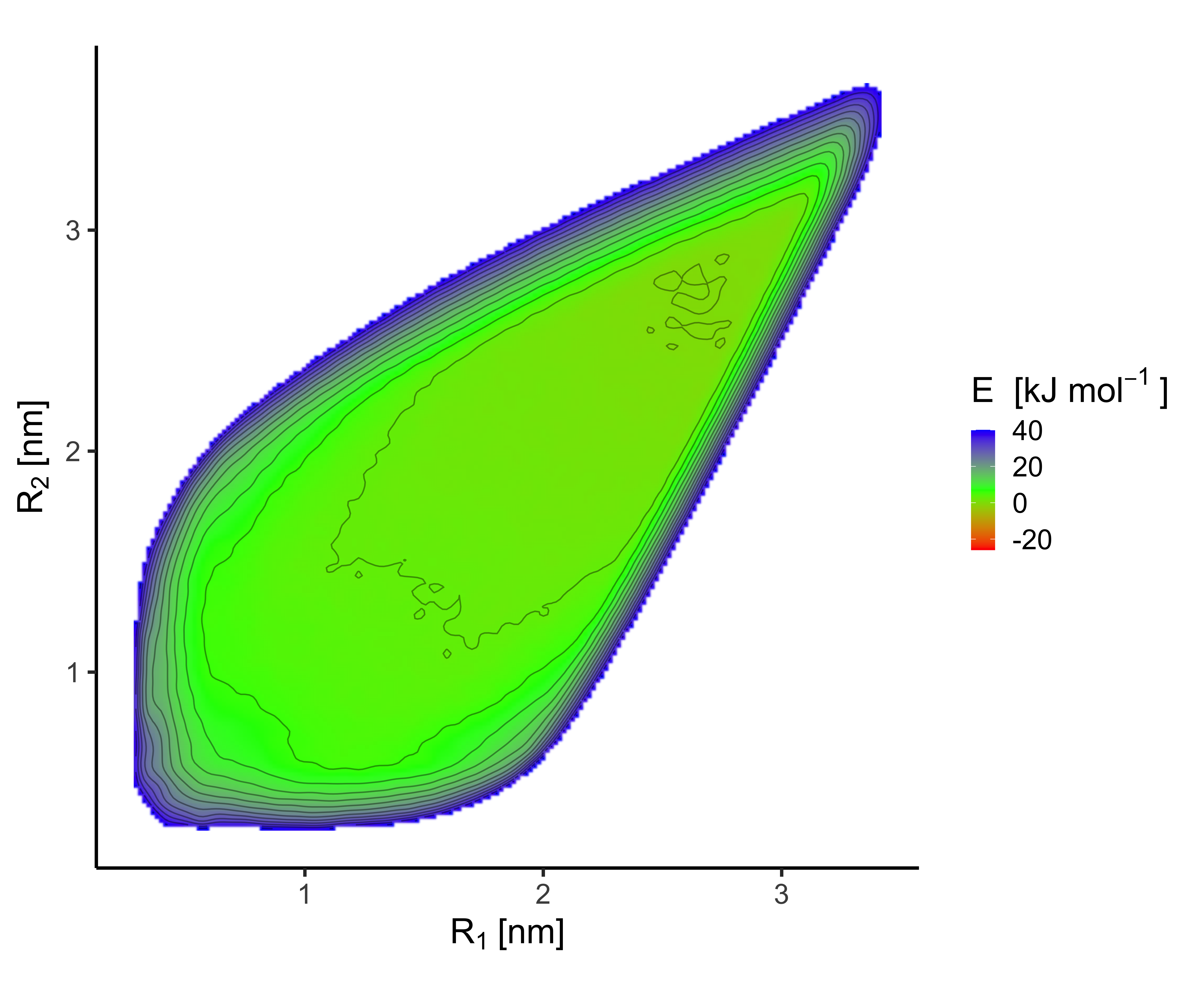}
 \caption{ Free Energy Surface for a Cl$^{-}$ ion and tri-ethylene glycol side chains.   }
   \label{fig:CVs}
\end{figure}

\vspace{1cm}
\begin{figure}[h!]
\center
  \includegraphics[width=12cm]{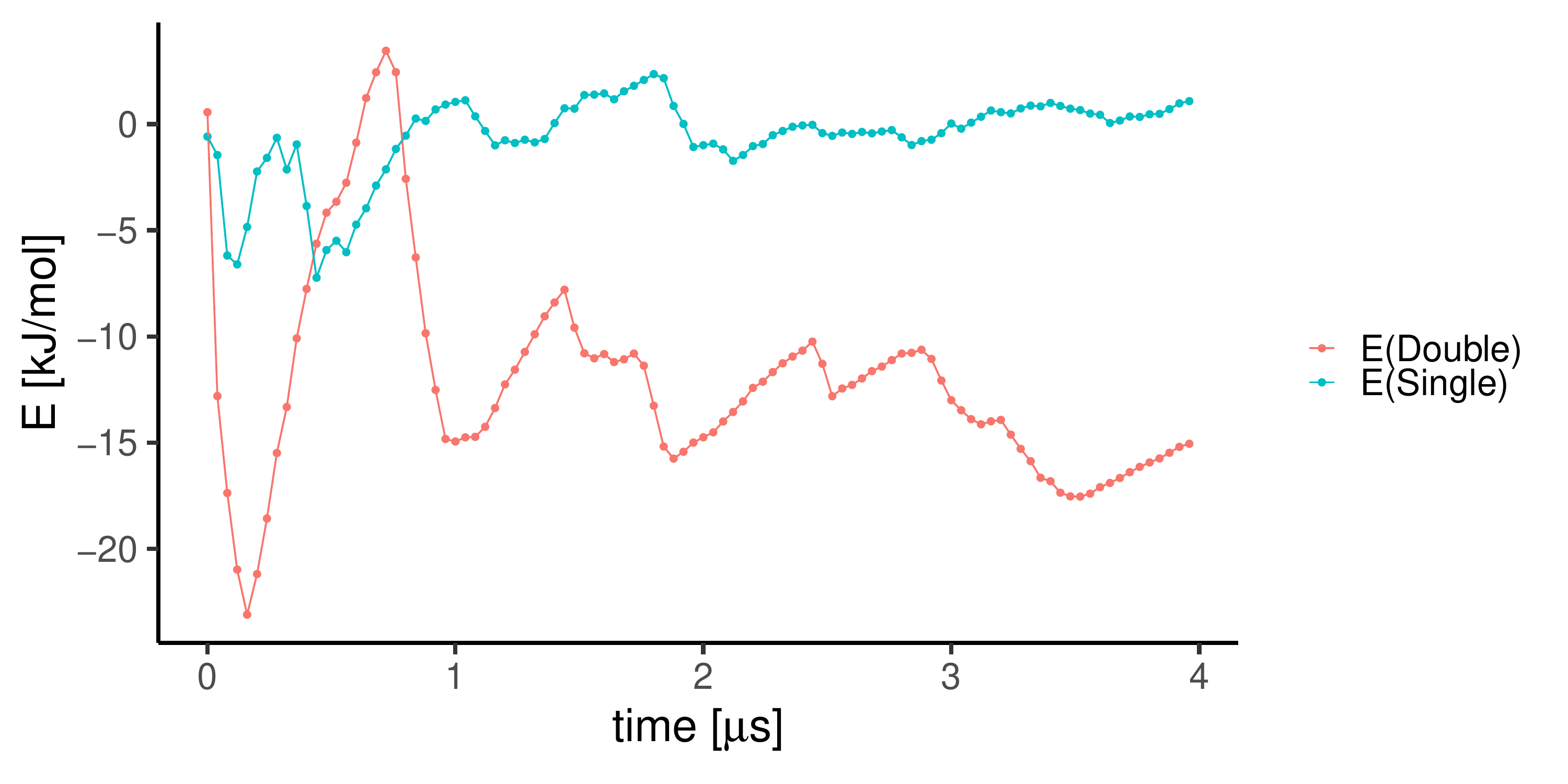}
 \caption{ Time dependance of the single and double binding energies for Na with pgT2.  }
   \label{fig:HILLS}
\end{figure}

\vspace{1cm}
\begin{figure}[h!]
\center
  \includegraphics[width=13cm]{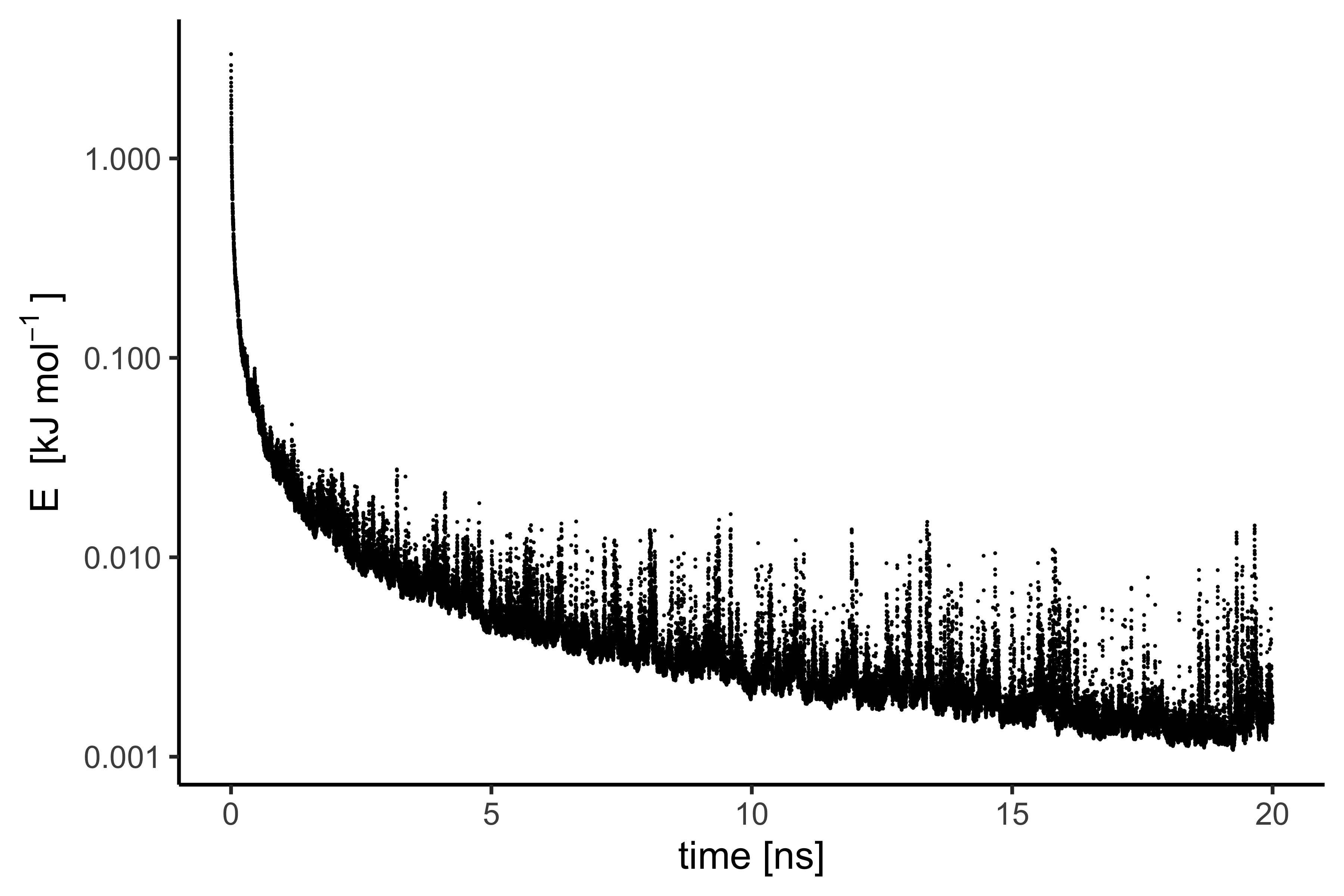}
 \caption{ Height of hills being deposited in the simulation, averaged across the 200 walkers.  }
   \label{fig:HILLS}
\end{figure}

\clearpage
{\footnotesize
\bibliographystyle{unsrt}
  \bibliography{Bib_File.bib}}